**Head-Independent Time-Invariant Infiltration Rate in Aquifer Recharge with Treated Municipal Wastewater**

Roy Elkayam*,†,‡ and Ovadia Lev‡

†Mekorot Water Company, Lincoln Street, Tel-Aviv – Yafo, 6492105, Israel.

‡The Institute of Chemistry, The Hebrew University of Jerusalem, Jerusalem, 9190401, Israel

E-mail: relkayam@mekorot.co.il

**Abstract**

Means to increase water resources are essential in regions grappling with water scarcity and growing populations. Soil aquifer treatment (SAT) is a cheap, low maintenance, low-energy method to supply water for irrigation of crops consumed raw or even for drinking purposes. However, the most expensive cost-component of SATs is the land use, the infiltration basins the area of which is inversely proportional to the infiltration rate, the most important characteristic of SAT basins design and operation, which until now was believed to be time-dependent and, therefore, difficult to predict. Focusing on the Shafdan SAT in Israel as a showcase and using a decade's worth of data from 50 recharge basins, we study the time dependence of the infiltration rates. The study reveals a noteworthy consistency in the decline of effluent levels during the drainage phase across various flooding events, signifying a constant, head-independent infiltration rate. 97% of over 40,000 flooding events showed this behavior. Furthermore, the infiltration rate calculated in this manner provides good predictions of the average infiltration rate during the entire wetting phase.

The water-level-independent infiltration rate is a general feature. It was found in all the 50 studied basins, regardless of the soil sand content, commissioning year, operation conditions and season. The constant infiltration rate law revealed in this study simplifies the prediction of the flooding cycle duration and will facilitate simplified predictive modeling of multiple basins SAT systems. Our research may extend beyond SAT systems, offering insights applicable to other managed aquifer recharge methods, crucial for effective water resource management, ensuring environmental compatibility.

This data-driven, large-scale study confirmed several conventions, usually accepted by water recharge practitioners but lacking statistical background. These include the seasonal variation of the infiltration rate, whereby the infiltration rate in the summer is much higher than in winter.

**Keywords:** Soil Aquifer Treatment (SAT), Recharge Basins, Infiltration Rate, Managed Artificial Recharge, Soil Infiltration

## Introduction

As cities in arid and semi-arid areas with growing populations face water shortages and frequent droughts, recycling wastewater transcends mere pollution control to an additional water resource (Grant et al., 2012; Schwabe et al., 2020). Soil Aquifer Treatment, SAT is long recognized as a reliable and sustainable means of effluent treatment, enabling wastewater recycling and safeguarding precious water resources with minimal use of manpower, energy, chemical consumption, and waste generation (Banin et al., 2002; Dillon, 2005; Dillon et al., 2019; Elkayam, 2019; Grinshpan et al., 2021; Kümmerer et al., 2018; Mansell & Drewes, 2004; Ying et al., 2003). The process involves flooding recharge basins with treated wastewater (TW), which then infiltrates into the aquifer. As it infiltrates through the unsaturated zone and flows through the saturated zone, contaminants are gradually removed and the TW is cleaned (Elkayam, Michail, et al., 2015; Elkayam, Sopliniak, et al., 2015; Grinshpan et al., 2021). The water is mostly reclaimed for reuse, and in some cases, it can be used for unlimited irrigation of crops eaten raw without additional treatment(Elkayam et al., 2018). The frequency and duration of the intermittent flooding of the recharge basins determine the infiltrated effluent capacity and, consequently, the size of the basins, which is the most costly resource in SATs. In addition to shedding light on effluent treatment mechanisms, the study of SAT systems offers insight into natural water recharge processes. In this context, it is notable that the Shafdan SAT system recharges over 100 meters of water annually (Barkay-Arbel et al., 2022), a significant figure given that Israel's average annual precipitation is about 0.6 meters (Hochman et al., 2020). Thus, the Shafdan system serves as a case study of artificial aquifer recharge, offering a unique opportunity to examine natural aquifer recharge phenomena on an accelerated scale. This becomes particularly significant in light of global warming, which induces extreme weather conditions and floods — challenges that can be alleviated by diverting excess water to artificial aquifer recharge systems. (Bergeson et al., 2022; Hossain-Anni et al., 2020; Li et al., 2022; Mahapatra et al., 2020).

The primary objective of this study is to conduct a meticulous analysis of the Infiltration rate (Ir) within an artificial recharge basin across a long specified time frame. To achieve this goal, we will utilize a comprehensive dataset spanning a decade, encompassing information from all the 50 regularly monitored recharge basins within the Shafdan SAT system.

Our central aim is to establish a compelling case that, for each individual basin during every flooding event, the infiltration rate remains constant throughout the entire drainage time and can accurately predict the average infiltration rate for the entire event. We demonstrate that during each flooding event, as drainage occurs, the decrease in effluent levels follows a linear pattern, suggesting a consistently constant soaking rate. However, variations in the infiltration rate can occur across different flooding events and infiltration basins. We assert that the dynamically evolving clogging layer on the topsoil of the recharge basins, which affects the seepage rate)Bouwer, 2002(, is responsible for the sustaining the constant infiltration rate.

*Infiltration rate.* The aquifer recharge by surface spreading involves three phases (Figure 1). Once the flooding phase begins, the water level in the recharge basin rises until it reaches a target level (alternatively, the flooding rate is kept equal to the infiltration rate to maintain a constant effluent level). Subsequently, the water feed to the basin is terminated, marking the end of the flooding phase and the beginning of the drainage phase. The water level decreases at the infiltration rate since there is no inflow to the basin. After the water level reaches zero, the drying phase begins, and the moisture level decreases at the topsoil. Once cycle and drying times are set, the average infiltration rate is proportional to the average water loading, i.e., the average water feed per surface area, and is therefore proportional to the capacity of the basin or SAT system. The average infiltration rate is therefore inversely proportional to the size of the basin required to treat a given effluent feed to the SAT. The synchronization of the beginning and ending of the floodings, and the distribution of the incoming flow from the Waste Water Treatment Plant (WWTP) among the basins, are based on the expected infiltration rates of all the basins. The infiltration rate through a basin may vary by almost an order of magnitude depending on ambient conditions and basin history. The extensive research dedicated to theoretical prediction of SAT infiltration rates is, therefore, fully justified. However, this stands in contrast to the limited amount of validation studies of predictive models based on large, basin-wide SAT systems.

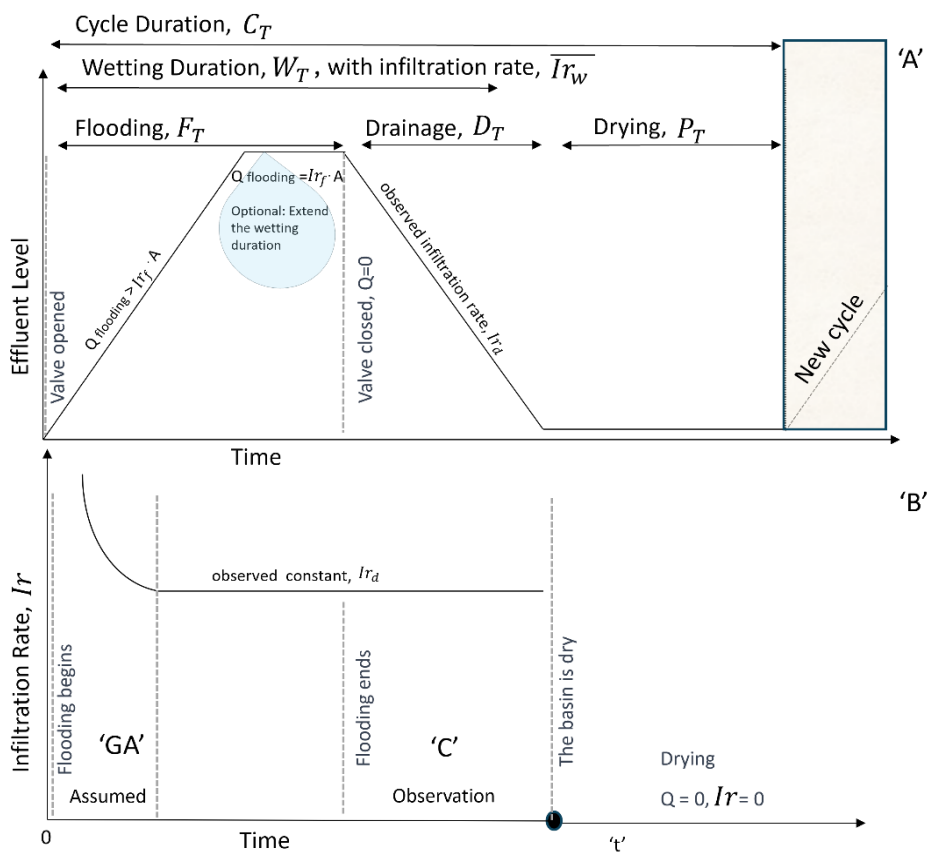

*Figure 1: A. Schematic operation regime for a recharge basin with the notations used in the article. The cycle comprises of three steps: (i) flooding, (ii) draining, and (iii) drying with $Ir_f$, $Ir_d$ infiltration rates. $\overline{Ir_w}$ is defined as the average infiltration rate during the wetting (i.e., flooding and draining) phase, $\overline{Ir_w} = \left(\int_0^{F_T}(Q/A)dt\right)/W_T$. The infiltration rates during flooding and drainage are also schematically shown in frame B. Three trends of the infiltration rate (Ir) are schematically depicted: 'GA' represents the Green and Ampt model (see text), 'C' represents a constant (head-indifferent) infiltration rate.*

The most comprehensive method to calculate the infiltration rate is by numerical solution of the non-linear Richards equation (Raats & Knight, 2018; Richards, 1931). However, this approach is time-consuming and more useful for the simulation of specific cases rather than formulating generalized rules on the infiltration rate through SAT spreading basins. Another useful approach is to calculate the infiltration rate using the Green and Ampt (GA) model (Green & Ampt, 1911) and its many modifications (Chen et al., 2019; Cui & Zhu, 2018). The GA model assumes a water saturation front propagating from the topsoil downward during the flooding and drainage phases. The water concentration front in the soil ranges between the soil water content prevailing below the front and the water saturation content. The depth of the water front is determined by the water balance, where the amount of effluent above ground and the added effluent down to the propagating front equal the amount of effluent that was pumped to the infiltration basin from the beginning of the flooding cycle divided by the basin's area (Green & Ampt, 1911; Tokunaga, 2020; Warrick et al., 2005). The driving force for the propagation of the saturated water zone downward is therefore the sum of the water head above ground (H), the water front depth below ground (L), and the capillary suction (h) immediately below the propagating front. As the flow is saturated, the infiltration rate (Ir) can be calculated by applying Darcy's equation (Bouwer, 1963, 2002; Lee et al., 2015):

Ampt, 1911; Tokunaga, 2020; Warrick et al., 2005). The driving force for the propagation of the saturated water zone downward is therefore the sum of the water head above ground (H), the water front depth below ground (L), and the capillary suction (h) immediately below the propagating front. As the flow is saturated, the infiltration rate (Ir) can be calculated by applying Darcy's equation (Bouwer, 1963, 2002; Lee et al., 2015):

$$Ir = K_s \frac{H + L + h}{L} \qquad (1)$$

$$Ir = K_s \frac{dH'}{dL} \qquad (2)$$

All the parameters in equations 1 and 2 are expressed in absolute values: *Ks* is the hydraulic conductivity at saturation, and *H'* is the pressure driving force (in meters). *L* is determined by the cumulative water feed per basin area from the beginning of the flooding cycle divided by the difference between water saturation and the initial water content. All the variables in equations 1 and 2 are time-dependent (the subscript, t, is omitted for simplicity). For a constant *Ks*, the GA model predicts that the infiltration rate will gradually decrease during the drainage phase since as *H* decreases, *L* increases to maintain water mass balance. Eventually, *Ir* should approach gravity drainage, where *Ir* equals *Ks*. The higher the water head above ground, the higher the infiltration rate (Furman et al., 2006; Philip, 1969). Several modifications of the GA model were proposed. The most notable modifications include multiple propagating fronts, each involving a (nearly) step change in water content. However, in all these modifications, when *Ks* and *h* are kept constant, the infiltration rate (*Ir*) increases as *H* increases. Adding to this complexity, entrapped gas impedes the infiltration rate (Mizrahi et al., 2016), making the prediction of *Ir* even more challenging.

Another approach to simplify the modeling is to use semi-empirical models where the infiltration rate, under constant head, is fitted by a time-dependent unction (Horton. R. E, 1940; Kostiakov A.N., 1932; Philip, 1969; Talsma & Parlange, 1972). These models involve 2 or 3 parameters that provide useful interpretation of ring infiltrometer tests. However, these approaches are less useful for predicting the duration of the infiltration cycle, as they only consider the dependence of *Ir* on time under constant head. Moreover, these models are usually compared based on short time scale, and only local infiltration rate is studied by the infiltrometers (Girei et al., 2019; Lake et al., 2009; Shukla et al., 2003). Finally, the models require two or three adjustable parameters, making real-life modeling, involving the effects of many ambient conditions on each fitted parameter excessively challenging.

An entirely different mechanism involves modeling the infiltration rate as a function of biological film formation at the topsoil and its evolution during the infiltration process. The model was thoroughly investigated by Herman Bouwer, a pioneer of managed water recharge and SAT research (Bouwer, 1991, 2002; Bouwer et al., 2014; Rice & Bouwer, 2013). Bouwer observed that the infiltration rate is often determined by a thin biological clogging layer, ranging in thickness from less than 1 mm to one cm (Bouwer, 2002). The infiltration rate can be calculated by the one dimensional integral form of Darcy's law.

$$Ir = K_f \Delta H'_f / L_f \qquad (3)$$

In this case, the hydraulic conductivity, $K_f$ and film thickness, $L_f$ depend on the cumulative infiltration rate from the beginning of the flooding cycle and the prevailing water head. As the film is deformable, $K_f$ is time and head-dependent.

In this study, we used data of over 40,000 effluent recharge cycles accumulated over a decade of operation of 50 Shafdan SAT basins. Our findings indicate that the water level decline during drainage follows a linear pattern (see Figure 1b), revealing a constant, head-independent infiltration rate. Furthermore, we demonstrate that the infiltration rate during drainage, $Ir_d$ can be used to predict the average infiltration rate during the entire wetting phase. This constitutes a crucial step towards data-driven modeling of the infiltration rate under variable ambient conditions and different water recharge operational histories. However, it is important to note that this does not imply that the infiltration rate does not vary between the different flooding events and across different seasons.

**Methods**

*Study Site Description.* The study area is the Shafdan SAT system, a large-scale wastewater reclamation plant (see Figure 2). The plant processes approximately 140 million cubic meters of wastewater annually, which flows from the Shafdan wastewater treatment plant (WWTP) and distributed to 70 infiltration basins covering a total area of 1.1 km$^2$ (Barkay-Arbel et al., 2022; Elkayam, Michail, et al., 2015). In this study,

We utilize a comprehensive dataset spanning a decade, which includes data from 50 out of the 70 recharge basins within the Shafdan SAT system. These 50 basins are continuously monitored using IoT devices, while the remaining 20 basins are operated manually without sophisticated monitoring.

More information on the Shafdan SAT is provided in the Supplementary Information (SI). Briefly, the first spreading lagoons in the Soreq-1 field (Figure 2) were constructed in 1977 and consisted of 19 recharge basins. An additional five multi-basin recharge fields (Yavne 1-4 and Soreq 2) were gradually commissioned in the next 15 years. The SAT was constructed in the rolling sand dunes above the Israeli coastal aquifer. Their upper layers comprise of fine sand underlain by low-density layer of less than 1 m due to frequent tillage. Like most SAT treatments in the world, the Shafdan SAT are located in sedimentary soil and their upper layer is mostly fine sand. Figure S1 in the SI shows that the lithology of the basins varies considerably. According to Mienis and Mualem (Mienis, 2013) it varies sometimes even within the same basin.

The lithology mostly comprises of layers of different mixtures of fine sand, silt, marl, and calcareous soil interrupted by thin clayey aquitards with varying levels. In some basins, the clayey layers are situated in the upper few meters, corresponding to the intermittently saturated layer described by the GA model (*vide supra*), while in others, these layers are found only at deeper locations. The lithological variability results in widely different infiltration rates ranging between less than 1 and over 10 cm/h, depending on the season and operational conditions. The widely different lithology and infiltration rates in the different basins underscores the robustness of the conclusions based on the Shafdan SAT studies.

*Data Collection.* The data source for this research was the Shafdan database, which has accumulated over four decades of operational knowledge regarding the Shafdan SAT reclamation systems. This dataset includes historical records of flowrates to the basins and the effluent levels in the basin, as well as operational parameters for 50 recharge basins. We utilized operational data from the recharge basins spanning ten years, focusing on time series data comprising the level, flowrate, and valve status for each time step within each recharge basin. The data underwent preprocessing to ensure quality and consistency, involving the identification and removal of anomalies or errors resulting in the rejection of 702 cases of anomalous data. More details and examples of rejected cycles are provided in the SI (Figure S2).

*Data Analysis.* A flooding event was defined as the time period between two consecutive openings of the inlet valve to an infiltration basin, corresponding to the initiation of the consecutive flooding cycles. Within each event, time series measurements of the effluent level were categorized as "Flooding", "Drainage", and "Drying", based on the status of the flooding valve and flow conditions. In the "Flooding" phase, the inlet valve to the basin is open, and the flow exceeds zero. The "Drainage" phase is characterized by closed inlet valve and non-zero effluent level in the basin, $H > 0$. The "Drying" phase was characterized by a closed inlet valve and the effluent level, $H$ equals 0. Accordingly, the wetting phase duration ($W_T$) equals the flooding duration ($F_T$) plus the drainage duration ($D_T$). The drying duration, denoted by subscript p, (for parched), completes the flooding cycle time ($C_T$), where $F_T + D_T = W_T$, and $C_T = W_T + P_T$.

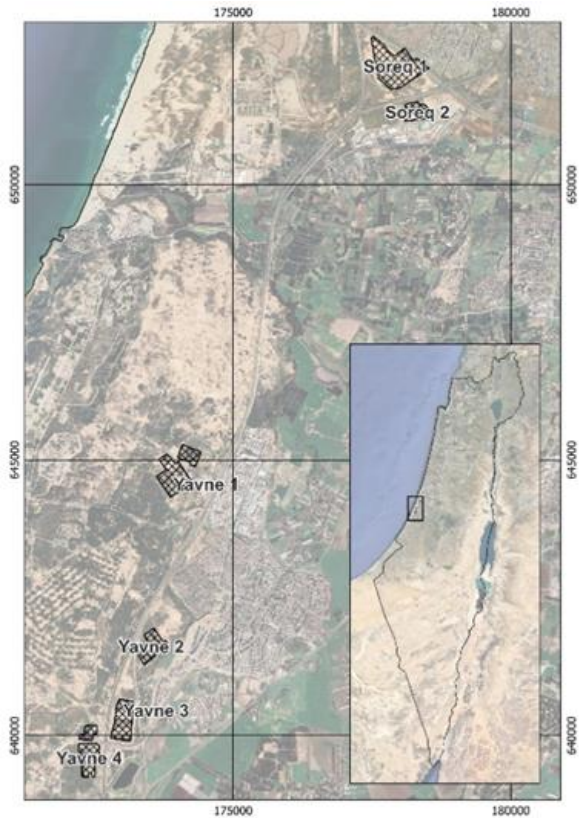

*Figure 2: Location Map of the Recharge Basins. Displays the geographical locations of the Shafdan recharge basins: Soreq 1, Soreq 2, Yavne 1, Yavne 2, Yavne 3, and Yavne 4. The inset map provides an overview of their locations in Israel.*

*Infiltration Rate during the Drainage Phase ($Ir_d$):* For each flooding event, the infiltration rate (in cm/h) was calculated based on the rate of decrease in water level during the basin drainage. The relative error in the measurement of H close to zero water level is excessively high due to several reasons: i) there is a small difference of the soil level at the two opposite sides of the basin; ii) the soil level is uneven due to tillage-induced ramps; and iii) there is some water level inaccuracy and imprecision due to instrumental errors. It should be noted that the level gauges are infrequently calibrated in the Shafdan SAT, and there is no record of inaccuracies found in the calibrations. Therefore, the calculations of the infiltration rate during the drainage phase was conducted only for H>14 cm. The instantaneous infiltration rate during drainage, $Ir_d$, was calculated based on the slope of the water level over time, starting from the initiation of the drainage phase. Linear regression analysis was applied to the dataset. A linear dependence of the water level on time with high Coefficient Of Determination, COD ($R^2$) values suggest a head-independent infiltration rate.

*Infiltration rate during the flooding phase ($Ir_f$):* In theory, it would have been equally straightforward to calculate the infiltration rate during the flooding phase by fitting the data to the linear expression in equation 5:

$$Ir_{f,t} = Q_t/A - (dH/dt)_t \tag{5}$$

Equation 5 represents a mass balance, similar to the method used to evaluate $Ir_d$, but it incorporates the non-zero, variable inlet flowrate to the basin ($Q_t$). The subscript, $t$ underscores the temporal nature of this mass balance. However, the accuracy of the prediction of $Ir_f$ depends on the quality and resolution of the flowrate *vs.* time dataset, which, for the Shafdan SAT, is poorer than the water level *vs.* time dataset, and, unfortunately, both dependencies are needed for $Ir_f$ determination by equation 5. Moreover, the acquisition of water level and inlet flowrate (i.e., $H_t$ and $Q_t$) is not synchronized in the Shafdan SAT, as these outputs were designed to fulfill different, unrelated functions. Therefore, $Q_t$ and $H_t$ refer to slightly different times. Finally, the flow to each of the infiltration basins is distributed from a central pipeline network, and therefore, opening or closing an inlet valve to one of the basins changes the flowrate to all other basins that are filled simultaneously.

Several examples of $Q_t$ during a few typical flooding cycles are delineated in Figure S3 in the SI to demonstrate how fast the flowrates to the basins change in time. More generally, Figure S4 in the SI shows several histograms of the relative standard deviations of $Q_{i,t}$ normalized to the average $Q_i$ at the $i_{th}$ flooding cycle. The distributions of the flowrates confirm and generalize the observations provided in Figure S2 and show that the flowrates vary considerably during the flooding cycles. The high intra-cycle variability of Q in Figures S3 and S4 demonstrates why a high-resolution time series is vital for the accurate evaluation of $Ir_f$ by equation 5.

Therefore, we had to resort to different approaches for the prediction of the average infiltration rate during the wetting phases based on $Ir_d$. We investigated whether $Ir_d$ (which is head-independent and therefore also time-independent) would enable us to predict the average infiltration rate during the entire wetting phases, $\overline{Ir_w}$. It is desirable to obtain a confirmation that $\overline{Ir_w}$ can be approximated by $Ir_d$. Note, that this requirement is much less stringent than the requirements that $Ir_f$ is time-independent and equal to $Ir_d$, which, theoretically, could be verified by equation 5. A second approach would be to set a basin-specific but season- and year-invariable correlation between $\overline{Ir_w}$ and $Ir_d$, (e.g., in the form of equation 6).

$$\overline{Ir_w} = k1 \cdot Ir_d + k2 \tag{6}$$

This would be sufficient to compute the wetting time ($W_T$) required to infiltrate a given hydraulic loading per cycle ($HL$) based on accurate prediction of one parameter, $Ir_d$. $HL$ is defined as the total effluent volume per area that is fed to a basin during a cycle divided by the cycle time. $HL$ is given by the integral of the flowrate to the basin ($Q_t$) over the flooding time ($F_T$) divided by the basin area ($A$) and cycle time, $C_T$,

$$HL = \frac{1}{C_T} \int_0^{t=F_T} \frac{Q_t}{A} dt = \overline{Ir_w} W_T / C_T \tag{7}$$

**Results**

*Outliers:* Throughout our analysis, we encountered several outliers in the basin operation data, mostly due to faulty meters and communication errors. Typical abnormal behaviors that could be classified as outliers are presented in Figure S2 in the SI. For example, instances such as 24-hour flooding periods that do not constitute proper flooding events due to rapid fluctuations of the water level, faults in the level measuring instrument, communication faults resulting in frozen values in the basin level measurements, insignificant level rise during the flooding events, and valve status faults that contradict the water level changes. For each drainage dataset corresponding to a specific flooding event, we took into account all descending sequences that lasted greater than 3 hours. Sequences that didn't comply with this constraint were removed. The number of removed cycles was 702, constituting 1.5% of the examined cycles. As for the flow measurements, although there were not many events, the deviations were significant, often reaching hundreds of percent of the physical value that could flow through a pipe of the feed diameter. Therefore, we smoothed the deviations by replacing each flow measurement that exceeded the 99.9th percentile with the previous value.

*Effluent Level Trends in Recharge Basins:* Figure 3 illustrates a sample of the effluent level changes over time in three recharge basins during different seasons. Each time step is categorized according to its operational status, namely, 'flooding,' 'drainage,' and 'drying' and assigned a different color. The figure demonstrates a consistent linear trend in the decline of effluent levels during drainage (in green). Visual examination of the decline of the water level in time revealed that the decline was always linear regardless of the season and operational conditions, such as the duration of flooding, the frequency of floodings, and the duration of drying periods between consecutive floodings. These observations prompted further investigation to confirm the hypothesis of a linear level decrease.

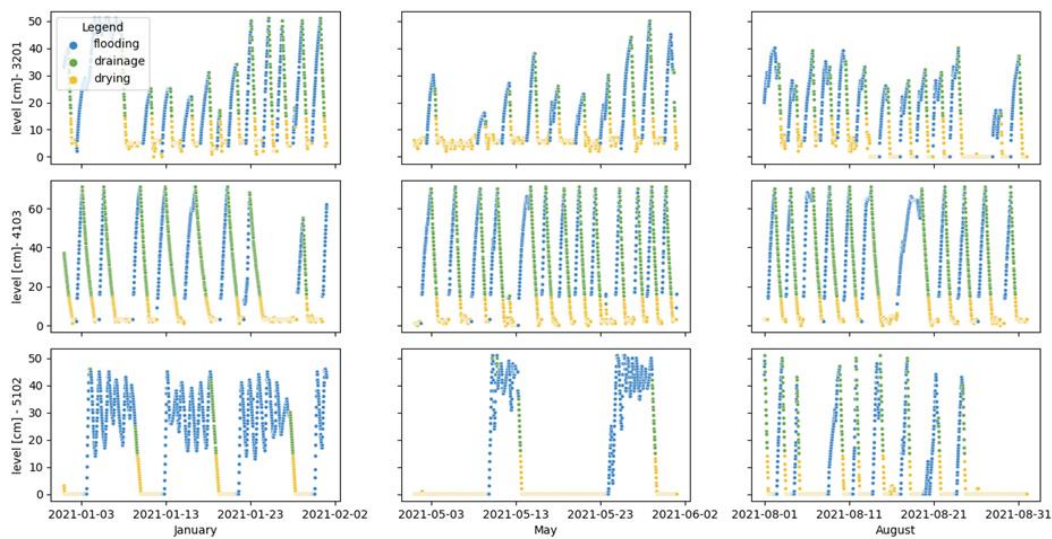

*Figure 3: Effluent level trends in three recharge basins. The effluent level trends over time in three recharge basins during January, May, and August are demonstrated. Operational statuses (flooding, drainage, drying) are classified at each time step, highlighting a consistent linear decline in effluent levels (shown in green) across varying operational conditions.*

*Infiltration Rate Analysis.* The comprehensive investigation of all flooding cycles from all infiltration basins, spanning ten years, yielded a total of 45,218 flooding events available for analysis after rejection of 702 abnormal cycles (1.5%). Within each of these events, data points classified under the 'Drainage' status were subjected to individual linear regression analysis. The results, as depicted in Figure 4, showed that linear effluent level decline, corresponding to time and head-independent infiltration rate, describe the data very well. 91% and 95% of the data cycles exhibited linear level decline with coefficient of determination, $R^2$ larger than 0.95 and 0.9, respectively. This is remarkably high, especially considering that there was no effort to screen out outlier data points (any descending sequence was acceptable), which are rather common in large-scale operations exposed to extreme ambient conditions. This is notable also since the raw data are solely used for the immediate control room operation and are not processed in the Shafdan afterward.

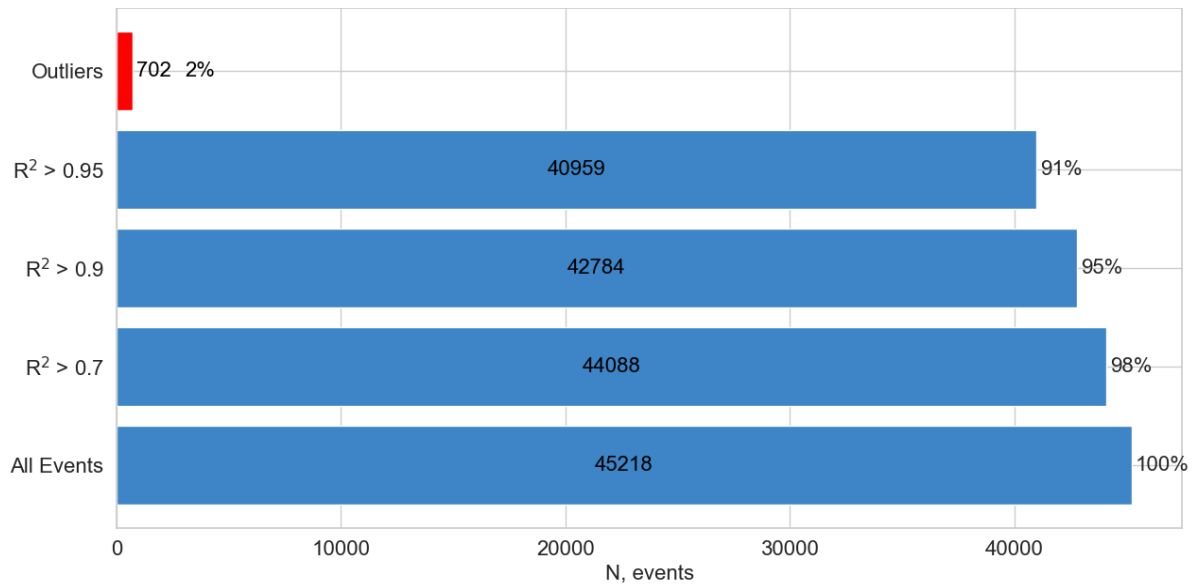

*Figure 4: Results of linear regression analyses conducted on the effluent level decline during the drainage phase. The data reveal a strong linear relationship, with 98% of the cycles displaying linear decrease of H, with $R^2$ values exceeding 0.7 and 91% surpassing 0.95.*

*Top-Down Mechanistic insight:* The fact that the head-dependent model fits the time-trace data so well does not necessarily imply that it is the best mechanistic model. True, the obtained fit is good enough to base predictive simulations of the infiltration rate on the linear decline rate, but the unprecedented availability of comprehensive field data provides good opportunity to obtain top-down mechanistic insight as well. Instead of postulating a mechanism, deriving a model and investigating its validity by field-observations, we followed an opposite approach by examining the best head-infiltration rate dependency. To that end, we investigated the logarithmic dependence in equation 8.

$$Log\ (Ir_{d,t})\ =\ C_1 + C_2\ Log\ (H_t) \qquad (8)$$

Where $C_1$ and $C_2$ are fitting constants, representing the coefficient and the power of a power function, $Ir_{d,t}=C_1 H_t^{C_2}$. *Note, that $C_2 = 0$ corresponds to a head-independent infiltration rate, $C_2 = 0.5$ corresponds to models in which potential energy (e.g., water head) is converted to kinetic energy (depending on the square of the velocity), and $C_2 = 1$ corresponds to models in which the flow through a constant hydraulic barrier takes place. The statistics of all the basins are presented in Table S1 in the SI, including the average, minimum, and maximum of $C_2$ obtained in the linear correlations of all the flooding cycles in the various basins. Figure 5 depicts the histogram of the average $C_2$ for each of the 50 basins. It is remarkable to note that the values of $C_2$ in all 42,000 studied cycles span a narrow range, -00039 (in basin 4201) > $C_2$ > -0.06 (in basin 6203). The average $C_2$ spans an even narrower range, -0.006 (in basin 4202) > $C_2$ > -0.02 (in basin 5202), confirming that the GA model and the intuitive positive dependence of the infiltration rate on the water level above the bottom of the basin failed to represent the showcase of 42,000 cycles in the 50 basins of Shafdan SAT. The negative values of all the computed pre-logarithmic coefficients, $C_2$ indicate that the infiltration rate depends on the depth of the impounded effluent, but, surprisingly, the dependence is opposite to the prediction of the GA model, higher water level correlates with (slightly) lower infiltration rate. Due to the large amount of data, despite the absolute low level of $C_2$, the negative sign is statistically significant (refer to Table S1 in the SI section). First, the average COD, $R^2$ of the linear correlations of all the cycles in all the basins was greater than 0.98. Secondly, the p-value for rejecting the null hypothesis of positive head dependency or truly head-independent rate (i.e., averaged $C_2 \geq 0$) in any of the basins is surprisingly small, less than $10^{-5}$, indicating that the negative correlation is not a coincidence. However, the negative effect is very small and holds no practical significance. For instance, an increase of 10 cm in water level reduces the infiltration rate, on average, by 2%, which is lower than the statistical margin of error of our level measurements.

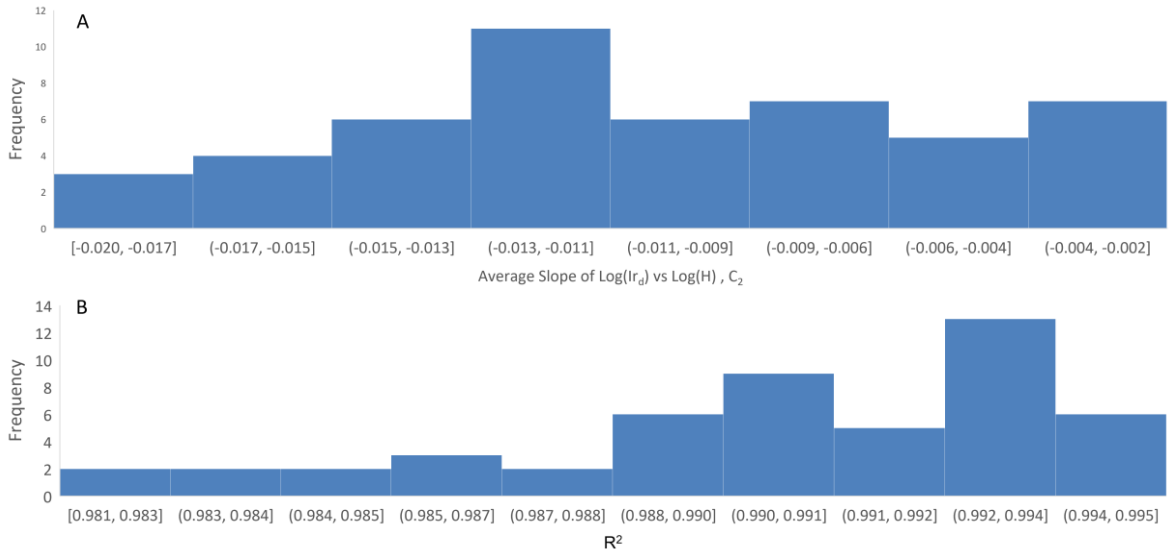

*Figure 5: Histograms of (A) the average coefficient, $C_2$ in equation 8 in all the studied infiltration basins, and (B) Histogram of the average COD for the 50 studied basins. All the values are >0.980 indicating strong logarithmic dependence, with very low dependence on the water head, H.*

*Seasonal variation of the infiltration rate:* As an example for the usefulness of the linear level decrease approximation, consider the seasonal variation of the infiltration rate during drainage. The fact that the infiltration rate is constant in each cycle allows for easy analysis of trends in the permeability of the lagoons during different seasons over a decade. The left frames in Figure 6 provide examples of the fitted $Ir_d$ variation over time, categorized by seasons. A close examination of the infiltration rate time-series in the figure reveals consistent patterns observed across most cases. Each season exhibits lower infiltration rates during the winter months, with $Ir_d$ gradually increasing, reaching peak infiltration rates during the summer season. Subsequently, there is a reduction in $Ir_d$ as winter returns. This cyclic pattern is evident in the left frames of Figure 6, highlighting the influence of seasonal weather fluctuations on infiltration rates and demonstrating the enduring impact of seasonal variation on infiltration rates in specific basins. However, when considering multi-year trends, such as the winter of 2017 in basin 3201 or the winter of 2019 in Basin 5102 (watermarked), it can be observed that within the same basin, values of the infiltration rate in the winter can be as high as those observed in the summer of other years (e.g., 2015, 2021). The right side of Figure 6 illustrates the distribution of $Ir_d$ values by season. Within this distribution, the summer season stands out with a distinct cluster of higher $Ir_d$ values, emphasizing the influence of meteorological factors on the infiltration rates. Similar observations showing that infiltration rate is higher in the summer were reported before (Jaynes, 1990; Lin et al., 2003), including reports on seasonal trends in the Shafdan SAT itself, though never before showing such consistency and based on such a large database. In addition, it's evident that while seasonality plays a significant role, other basin-specific factors contribute to the observed variability. Figure 6 reveals a large range in $Ir_d$ values across the recharge basins, spanning from 1 to 8 cm/h, indicating significant variability in the infiltration rates. These variations highlight the complex interplay between seasonal weather patterns, basin characteristics, and operational factors in determining infiltration rates.

*Prediction of $\overline{Ir_w}$ by $Ir_d$.* It could be anticipated that if the infiltration rate is time- and head-independent during the drainage phase, it would be likely be head-independent for most of the wetting phase as well. Therefore, $\overline{Ir_w}$ should be roughly equal to $Ir_d$ or at least highly correlated with it, which would provide means to estimate $\overline{Ir_w}$ based on $Ir_d$. Figure 7 presents the time trace of $Ir_d$ (in blue dots) and $\overline{Ir_w}$ (in pink) in 9 basins, qualitatively confirming that the infiltration rates in the drainage and the wetting phases in each cycle are indeed close to each other. It can be observed that the differences in magntude between the blue and pink dots are considerably smaller than the value of $Ir_d$. Therefore, we verified the validity of the relationship

$$\overline{Ir_w} = k \cdot Ir_d \tag{9}$$

where $k$ is a proportionality constant that ideally should be close to 1.

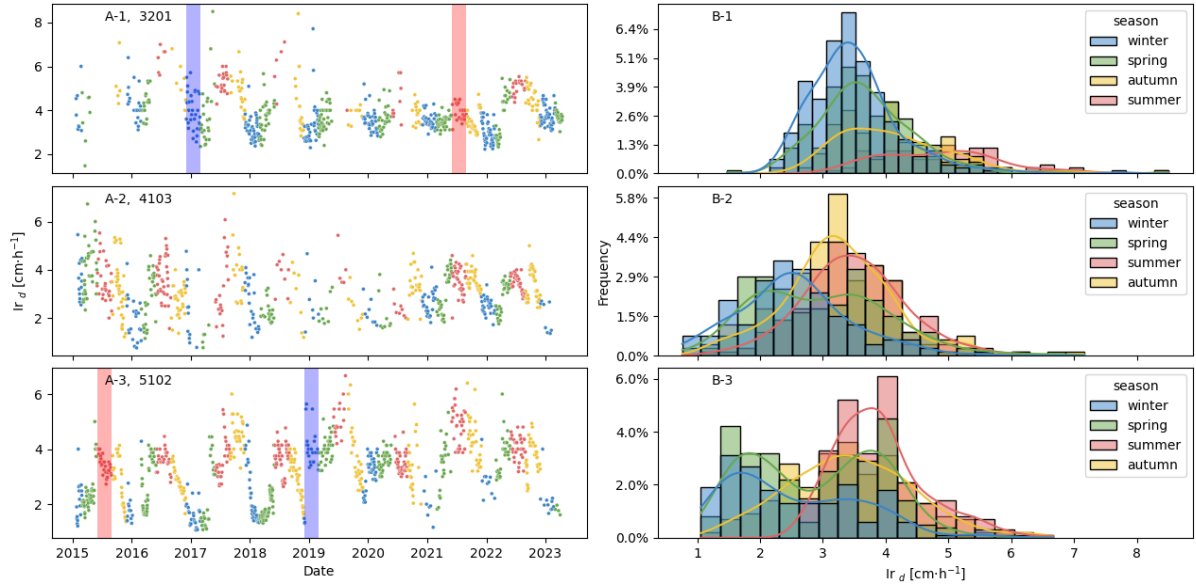

*Figure 6: Seasonal variation of the infiltration rate, showcasing the dynamic range of the Ir values over time. A1-A3: Distribution of Ir values as a function of time. Blue watermarks highlight the winter seasons of the years 2017 and 2019 in basin 3201 and 5102, respectively. B1-B3: Presentations of the distribution of infiltration rates during the different seasons (the symbol colors fit the colors of the bars in A1-A3). The average infiltration rate in the summer is highest and in the winter it is lowest, but there is overlap between the distributions.*

The accuracy of equation 9 was evaluated by examining the ratio of $Ir_d$ to $\overline{Ir_w}$ for all flooding events across all recharge basins. $\overline{Ir_w}$ was calculated from the actual water load during each cycle ($\overline{Ir_w} = \frac{1}{W_T}\int_0^{t=F_T}(Q/A)\,dt$). Given that the relative errors in flowrate measurements are much larger than in water level measurements, we employed an additional technique for outliers' removal at this stage. However, we endeavored to avoid introducing bias by not removing an excessive number of flooding events and maintaining statistically valid rules for the assignment of outliers. First, we checked whether the ratio between $\overline{Ir_w}$ and $Ir_d$ was normally distributed by the Shapiro-Wilk normal distribution test (Shapiro & Wilk, 1965). In all the basins, (except for basins 7302 and 7303) the Shapiro-Wilk test revealed that it is statistically justified to reject the null hypothesis that the population of the infiltration rate ratios is taken from a normally distributed population (last columns of Table S2 in the SI). Therefore, the Tukey's interquartile range method (Beyer, 1981; Vinutha et al., 2018) for outliers rejection, which is applicable for populations that are not normally distributed, was used. According to this method, the interquartile range (*IQR*) between the first (*Q1*) and third (*Q3*) quartiles is used to set outlier thresholds (fences). Data points falling above *Q3 + 1.5·IQR* or below *Q1 - 1.5·IQR* are classified as outliers. These criteria led to the exclusion of almost 9% of the data points. Detailed information on the removal of outliers in each recharge basin can be found in Table S2 of the SI. Figure 8 displays the linear correlations between $\overline{Ir_w}$ and $Ir_d$ in 9 basins. The dashed correlation lines were constrained to pass through the origin (as in equation 9) and the dotted lines represent the unconstrained correlations. A table depicting all the slopes of the linear correlations and the corresponding coefficients of determination, along with additional information about the number of outliers, are delineated in Table S2 in the SI. The average slope, k of the constrained correlation was 0.97, with an average relative standard deviation of 0.11, remarkably close to the expected identity slope, particularly considering the large variability of $Ir_d$ (refer e.g., to Figure 6). The 50 basins exhibited a good correlation between $\overline{Ir_w}$ and $Ir_d$, with an average $R^2$ value higher than 0.8, and a standard deviation of 0.09. Only one basin exhibited an $R^2 < 0.6$. Histograms of the average observed slopes and $R^2$ for each basin are depicted in Figure S5 in the SI. The highest $R^2$ encountered in any basin was 0.93, observed in basin 3203 in the Soreq field. It is notable, that for this basin, the outlier removal accounted for only 1%. Basins with lower correlation values were either in older fields that tend to have less accurate flowmeters or in basins where a high percentage of outliers removed, indicating poorer data quality.

Figure 7 shows that in certain basins, such as basin 7101, the $Ir_d$ values are higher than the $\overline{Ir_w}$ values, corresponding to $k>1$, while in others, such as basin 6303, the opposite trend is observed. We observed that in only 16 recharge basins, the $\overline{Ir_w}$ was larger than the $Ir_d$, suggesting that $Ir_f$ should also be larger than $Ir_d$. This observation implies that the infiltration rate decreases with time during the wetting phases, as expected based on GA, and Horton models, which predict decreased infiltration rate with time (Green & Ampt, 1911; Horton. R. E, 1940). However, in 34 recharge basins, the $Ir_d$ was larger than the $\overline{Ir_w}$, suggesting that $Ir_f$ was lower than the $Ir_d$. Chen and his team (Arye et al., 2011; Nadav et al., 2012) anticipated that, unlike freshwater, treated wastewater infiltration rate should increase with time, due to the higher water repellency of the hydrophobic biological film on the topsoil. The increased infiltration rate with time in water repellent soils was documented also in other sites with impounded TW (Arye et al., 2011; Feng et al., 2001; Magesan et al., 1999). However, the observation that, on the average, $Ir_d$ is larger than the $\overline{Ir_w}$ is not significant statistically, the standard deviation of k in the 50 basins is 0.11, much larger than the deviation of the average of k (0.97) from 1.

*Basin-specific correlations of $\overline{Ir_w}$ based on $Ir_d$*: In order to improve the prediction of $\overline{Ir_w}$ based on $Ir_d$, we examined the linear correlations between the two variables without imposing the pass-through-the-origin constraint. However, this analysis failed to improve the correlations significantly. The distinction between the dotted and dashed lines in figure 8 is minimal in most cases. Table S2 in the supplementary materials presents the complete data for the 50 examined basins which further underscore this observation. The average COD for the two parameter fit was 0.83, only marginally improved over the constrained fit (0.81). We attribute the very small intercept, averaging only 0.08 ± 0.26 cm/h, to systematic errors that are proportionate to $Ir$, such as inaccuracies in estimating the basins areas and inlet flowrates. These errors tend to have little or no impact on the deviation from $Ir_d$ values close to zero. The analysis of mean square error in Table S2 allows us to glean insight into the quality of the prediction of average $Ir$ throughout the wetting section. It is possible to calculate a measure for the average relative error by taking the square root of the mean square error, MSE divided by the mean infiltration rate in each basin (indicated as srMSE in Table S2), the average srMSE for all 50 basins is only 16%, and can be as low as 9% in cases where the data quality is very good, such as in basins 3203 and 3204 in the Soreq field. It is notable, that the constrained and unconstrained correlation exhibited the same relative srMSE, when written with two figures.

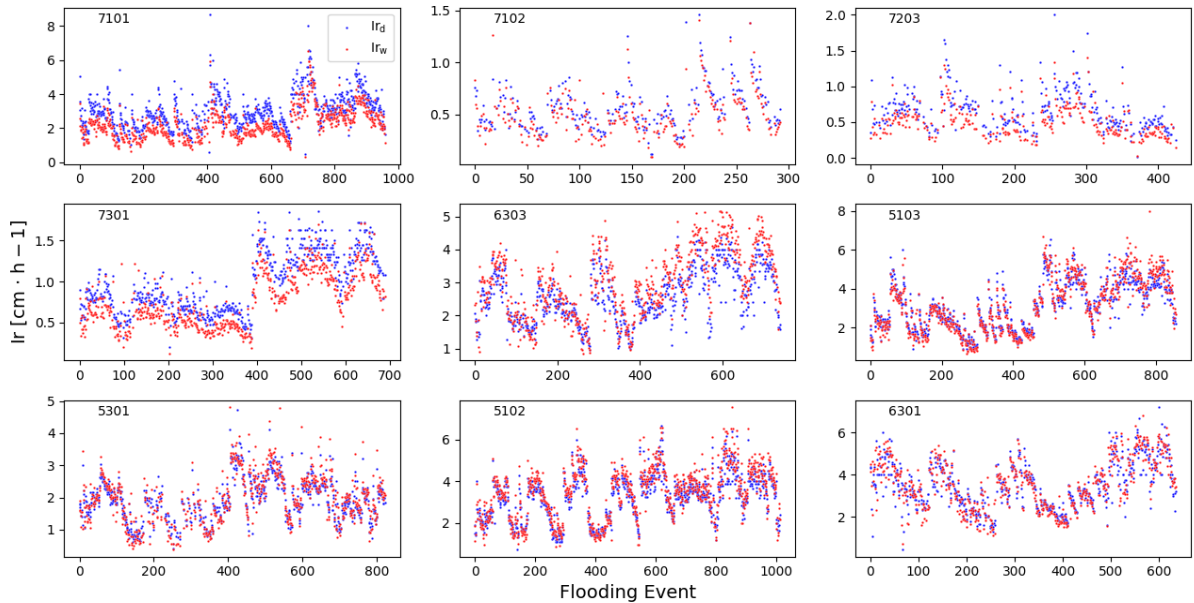

*Figure 7: Time trace of $Ir_d$ (blue) and $Ir_w$ (pink) in the last decade in 9 infiltration basins.*

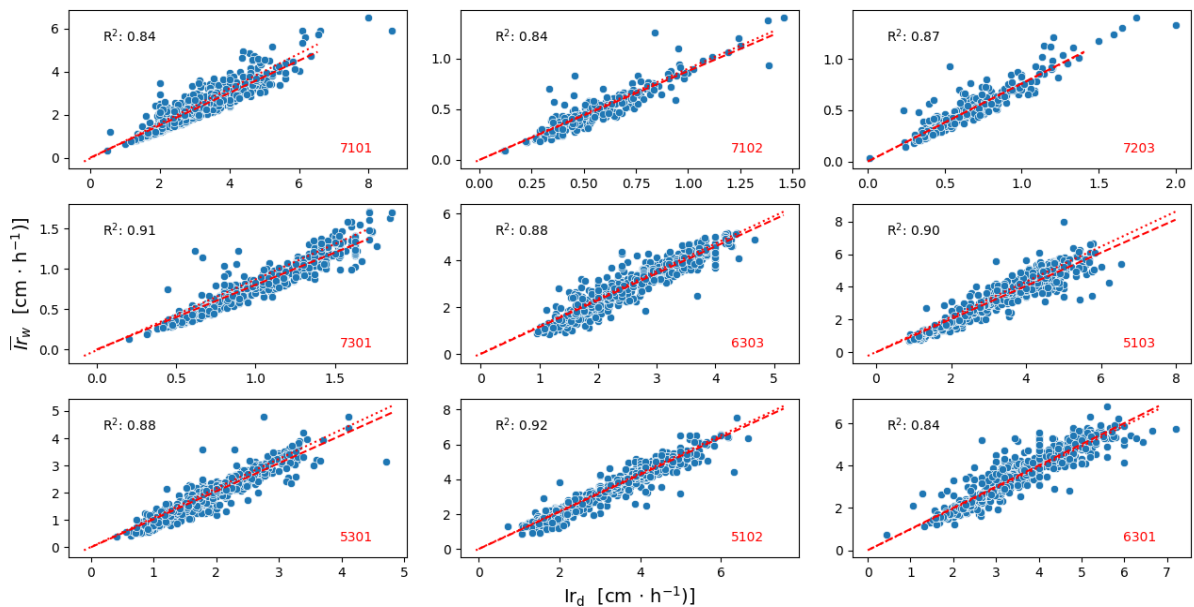

*Figure 8: Linear correlations between the average infiltration rate during the wetting phases, $\overline{Ir_w}$ and the head-independent infiltration rates during the drainage phase, $Ir_d$. The dashed trend lines were constrained to pass through the origin. The dotted lines represented unconstrained linear*

*correlations. In many cases, the two lines overlap. The $R^2$ value presented on the figure indicates a strong correlation between $\overline{Ir_w}$ and $Ir_d$.*

**Discussion**

The research provides a large-scale empirical evidence supporting linear decline in water levels during the drainage phase in recharge basins, confirming head-independent infiltration rate. The approach taken is data-driven, and the evidence is derived from a comprehensive dataset spanning ten years, 50 different recharge basins, and over 40,000 flooding cycles. This large-scale analysis is incomparable in magnitude to customary infiltration rate studies, which focus on one or a few sites, sometimes within a single basin (mostly by infiltrometer studies) or on a few basins in a single season (Barry et al., 2017; Goren et al., 2014; Haaken et al., 2016; Nadav et al., 2012; Racz et al., 2012). The vast database provides generality and robustness, despite having a larger scatter of the data points. While it is widely recognized that the infiltration rates are influenced by the water level (Horton. R. E, 1940; Kostiakov, 1932; Philip, 1969; Talsma & Parlange, 1972), the study reveals an almost head-independent infiltration rate. The linear correlation studies of the rates of water level decline were confirmed by a through regression analysis of $\log(Ir)$ against $\log(H)$. We observed a statistically significant (p-value ~ 0) negative correlation between the impounding water level in all recharge basins and the infiltration rate during drainage. However, we believe that the magnitude of the pre-logarithmic coefficient is so low that the negative dependency is only a theoretical issue and has little or no practical significance.

Furthermore, we demonstrated that the average infiltration rate throughout the wetting phase, and for that matter, also throughout the wetting cycle, can be computed by a basin-specific correlation between the average infiltration rate and the observed infiltration rate during drainage. The water loading (per square meter), the most important characteristic of an infiltration basin, can be easily computed based on the average infiltration rate during the drainage phase. Several issues regarding the generality of the conclusions, the compatibility, qualitative mechanistic explanation, and the significance of the conclusion warrant a brief additional discussion.

*Applicability*: The applicability domain of the research results pertains only to surface spreading SATs operated in flooding-drying cycles, and it is restricted to shallow, <90 cm SATs operating in a few-day flooding-drying cycles. However, the research results shed light on the mechanisms involved in infiltration processes and may offer insights into related processes, such as surface spreading lagoons operating under long cycles with a constant effluent level or managed aquifer recharge for stormwater capture.

*Generality.* Most SATs in the world are installed in sedimentary soils with fine sand over-layer, (Brito et al., 2009; Maliva, 2020; Page et al., 2018; Zheng et al., 2021). The showcase study examined nearly ten years of operation. It included 50 surface-spreading basins with vadose-zone lithology that varied considerably. The studied basins were constructed over a long period, 20 – 35 years before the beginning of the studied time frame. Finally, the infiltration rates also span a very large range, 1-10 cm/h, and even the average infiltration range in the basins varied considerable (see Figures 6,7). Despite this variability, all the basins followed the same trends irrespective of the soil sand and clay content, season, dry or wet years, commissioning year, and operation conditions including tillage. This wide range of attributes and the robust head-independent observation attest to the generality of the observed trends and research conclusions. Moreover, the research suggests that agreed concepts regarding infiltration rate *vs* head in managed aquifer recharge, including freshwater storage and storm-water capture and recharge, should not be taken for granted in view of the head-independent, time-invariant infiltration rate observed in this research.

Prospects for a predictive modelling of the infiltration rate: Two unexpected and related aspects that were found in this research offer promise for predictive modeling of the infiltration rate: i) the infiltration rate remains constant during the entire drainage phase, and ii) the existence of basin-specific correlation between the infiltration rate during drainage and the average infiltration rate throughout the cycle (equation 9). Specific predictive equations with high COD values for each basin are introduced in columns 11 and 12 of Table S2 in the SI. The average relative error in prediction of $\overline{Ir_w}$ by $Ir_d$ is only 16%. Moreover, it was observed that the average infiltration rate during the entire wetting phase closely matches the drainage infiltration rate, differing, on average, by only 3% (i.e., $\overline{Ir_w} = Ir_d$, corresponding to $k=1$ in equation 9). This difference is very small compared to the large variability of the observed infiltration rates within the same basin. Therefore, unlike previously proposed models for the infiltration rate (e.g., Furman et al., 2006; Horton. R. E, 1940; Kostiakov A.N., 1932; Sihag, 2018), a single parameter is sufficient to describe the infiltration rate, which simplifies prospective modeling.

*Compliance of the head – independent infiltration rate with hydraulic principles*: The time-invariant, head-independent infiltration rate during drainage deviates from previous infiltration model predictions. The Horton model (Horton, 1940) and the modified Kostiakov model (Sihag et al., 2017) predict a constant infiltration rate after prolonged drainage at constant head. The GA model (Green and Ampt, 1911) predicts that as H deceases in equation 2, L will increase due to water balance, and eventually, the infiltration rate will decrease and gradually approach the hydraulic conductivity, Ks. In contrast, the constant *Ir* in our study pertains to the entire drainage phase, including the highest water levels.

Mathematical modelling of the head–independent infiltration rate is beyond the scope of this data-driven analysis. However, it is easy to formulate a two - resistance model, by combining the GA model (Eq. 1) and the clogging film over-layer (equation 3). Adding the two hydraulic resistances in series, under a total pressure head, $(H+L+L_f+h)$, with the same water flux, *Ir*, passing through the two resistances, and assuming that the thickness of the film, $L_f$ is much smaller than the depth of the propagating water front, yields the relationship expressed in equation 4,

$$Ir = \frac{(L + H + h)}{\frac{L}{K_S} + \frac{L_f}{K_f}} \quad (4)$$

The total driving head ($L+H+h$) in the numerator increases from the beginning of the cycle. The denominator also increases, but in a more complex manner than in the GA model. In the GA model (equation 1), the resistance increases as the water saturation front travels downward and L increases. The denominator grows even faster than the numerator, and eventually, as $H$ decreases, the infiltration rate decreases and eventually approaches $Ks$. The denominator in the two-resistance model incorporates an additional non-linear contribution of the hydraulic resistance of the clogging film ($L_f/K_f$). The clogging film thickness increases with the cumulative infiltrated effluent passing through it due to biological growth and accumulation of suspended matter in and on the film.

The hydraulic conductivity of the film, $K_f$ is affected by conformational changes. Since the film is partly elastic, its hydraulic conductivity depends on the applied pressure. The compressing pressure constitutes only a fraction of the total pressure drop ($H+L+h$) over the saturated soil, and is practically always smaller than $H$. Thus, the resistivity of the film increases at the beginning of the cycle due to compression, deformation of the biofilm, and flattening of the biological material. However, it then reaches a maximum, and as $H$ starts to decrease, the resistivity decreases accordingly. In this phase, i) the portion of the total pressure drop on the growing saturated zone ($L$) is increased, ii) the semi-elastic film expands, and iii) there is a plausible release of entrapped suspended solids from the film. All three factors together compensate for the decrease of the water level and contribute to achieving the observed constant infiltration rate. Thus, surprisingly, the elasticity-induced, auto-regulation that is commonly found in water flow restrictors, arterial blood flow, pulmonology, etc., regulates the infiltration rate in SAT soils as well.

Examination of the time trends presented in this research (e.g., in figure 6 and 7) and the postulated mechanism responsible for the head-independent infiltration rate underscores the importance of the biological control of the infiltration rate. Only a responsive biological system (presumably, the clogging film) can allow constant infiltration rate during a cycle with such wide variability of intra-basin and inter-basin infiltration rates. This also underscores the importance of the qualitative understanding of the constraining action of the clogging biofilm throughout the SAT operation.

*Infiltration Dynamics*: Our findings demonstrate diverse trends of $\overline{Ir_w}$ vs $Ir_d$, indicating the simultaneous presence of two distinct mechanistic models within the same system: i) forecasts from the Green & Ampt (1911) and Horton (1940) models, suggesting a decline in infiltration rate over time; and ii) infiltration rates increase over time in water-repellent soils infiltrated with treated wastewater as predicted by Chen and colleagues (Arye et al., 2011; Nadav et al., 2012). Because of the frequent cycle time (i.e., a few days rather than weeks), we were able to identify these coexisting models before other mechanisms, such as suspended solids accumulation, emerged (Bancolé et al., 2004; Katznelson, 1989; Rice & Rice, 2013). In this context, we can explain the coexistence of these models by the semi-elastic film auto-regulation, which, in some cases, drives the system to behave more like the first model and in other cases like the second model. In our results, we observe the entire spectrum (Table 2 in the SI), ranging from basins where we documented only 1% of flooding events with $\overline{Ir_w}$ higher than $Ir_d$ over ten years (7301) to basins where 97% of flooding events exhibited $\overline{Ir_w}$ higher than $Ir_d$ over the same period (3204). Gleaning deeper insight into why different basins behave differently should involve incorporating many more parameters that influence the infiltration rate into a predictive model. The revelation that the infiltration rate is constant through the drainage phase and can be approximated by $Ir_d$ throughout the entire flooding cycle marks an important step towards developing such a model.

**Conclusions**

This study carries significant methodological and fundamental implications. It underscores the critical role of data-driven analysis in comprehending complex surface-spreading water recharge systems. Through data-driven analysis, patterns can be discerned even within fluctuating, noisy systems, such as soil aquifer treatment with its distributed basins and high operational and weather variability. This enabled us to confidently identify the head-independent infiltration rate law. The observation that only one parameter, $Ir_d$, remains constant during the drainage phase challenges traditional infiltration models.

The realization that only one parameter has to be predicted opens the door for machine learning to analyze the dependence of the infiltration rate on past and prevailing ambient and operational conditions. It is now just a matter of time before machine learning tools can be applied to accurately predict SAT performance.

Future research should aim to meticulously dissect meteorological and operational conditions to identify the factors that promote the formation of a less resistive thin layer, thereby maximizing infiltration rates. This investigation holds the key to unlocking a deeper understanding of the dynamics governing infiltration rates across various flooding events and basins.

he implications of this research may extend beyond artificial treated wastewater recharge methods to encompass Managed Aquifer Recharge (MAR) systems and Flood Management strategies. By directly measuring infiltration rates, we can enhance the management of recharge systems, ensuring their long-term effectiveness and environmental compatibility. As a result, this research can have broader significance in addressing global water resource challenges.

**Acknowledgement**

The authors are grateful for the financial support of Mekorot, Water Company Ltd and The Hebrew University.

**Supporting Information**

Supporting information includes lithological information of the Shafdan SAT lagoons, and tables and figures providing more statistical data.

**References**

Arye, G., Tarchitzky, J., & Chen, Y. (2011). Treated wastewater effects on water repellency and soil hydraulic properties of soil aquifer treatment infiltration basins. *Journal of Hydrology*, *397*(1–2), 136–145. https://doi.org/10.1016/j.jhydrol.2010.11.046

Bancolé, A., Brissaud, F., & Gnagne, T. (2004). Oxidation processes and clogging in intermittent unsaturated infiltration. *Water Science and Technology*, *48*(11–12), 139–146.

Banin, A., Lin, C., Eshel, G., Roehl, K. E., Negev, I., Greenwald, D., Shachar, Y., & Yablekovitch, Y. (2002). Geochemical processes in recharge basin soils used for municipal effluents reclamation by the soil-aquifer treatment (SAT) system. *Management of Aquifer Recharge for Sustainability. Proc. 4th Int. Symp. of Artifical Recharge (ISAR4). AA Blakema, Rotterdam, The Netherlands*, 327–332.

Barkay-Arbel, Y., Kohen, E., Megidish, E., Nadler, D., & Amran, S. (2022). *"Mey Ezor Dan" Agricultural Cooperative Water Society Ltd. Dan Region Wastewater Project Soreq Mechanical Biological Wastewater Treatment Plant Operation - 2021*.

Barry, K., Vanderzalm, J., Miotlinski, K., & Dillon, P. (2017). Assessing the Impact of Recycled Water Quality and Clogging on Infiltration Rates at A Pioneering Soil Aquifer Treatment (SAT) Site in Alice Springs, Northern Territory (NT), Australia. *Water*, *9*(3), 179. https://doi.org/10.3390/w9030179

Bergeson, C. B., Martin, K. L., Doll, B., & Cutts, B. B. (2022). Soil infiltration rates are underestimated by models in an urban watershed in central North Carolina, USA. *Journal of Environmental Management*, *313*, 115004. https://doi.org/10.1016/j.jenvman.2022.115004

Beyer, H. (1981). Tukey, John W.: Exploratory Data Analysis. Addison-Wesley Publishing Company Reading, Mass. — Menlo Park, Cal., London, Amsterdam, Don Mills, Ontario, Sydney 1977, XVI, 688 S. *Biometrical Journal*, *23*(4), 413–414. https://doi.org/10.1002/bimj.4710230408

Bouwer, H. (1963). Theoretical effect of unequal water levels on the infiltration rate determined with buffered cylinder infiltrometers. *Journal of Hydrology*, *1*(1), 29–34. https://doi.org/10.1016/0022-1694(63)90030-1

Bouwer, H. (1991). Role of groundwater recharge in treatment and storage of wastewater for reuse. *Water Science and Technology*, *24*(9), 295–302.

Bouwer, H. (2002). Artificial recharge of groundwater: Hydrogeology and engineering. *Hydrogeology Journal*, *10*, 121–142. https://doi.org/10.1007/s10040-001-0182-4

Bouwer, H., Lance, J. C., & Riggs, M. S. (2014). *All use subject to JSTOR Terms and Conditions II : Water land treatment of the aspects quality and economic Flushing Meadows project*. *46*(5), 844–859.

Brito, Gracieli, L. M., Schuster, Hans, D., Srinivasan, & Vajapeyam, S. (2009). Estimation of Annual Ground Water Recharge in The Sedimentary Basin of The River Peixe, Paraíba, Brazil. In *Advances in Water Resources and Hydraulic Engineering* (pp. 269–274). Springer Berlin Heidelberg. https://doi.org/10.1007/978-3-540-89465-0_50

Chen, S., Mao, X., & Wang, C. (2019). A Modified Green-Ampt Model and Parameter Determination for Water Infiltration in Fine-textured Soil with Coarse Interlayer. *Water*, *11*(4), 787. https://doi.org/10.3390/w11040787

Cui, G., & Zhu, J. (2018). Infiltration Model Based on Traveling Characteristics of Wetting Front. *Soil Science Society of America Journal*, *82*(1), 45–55. https://doi.org/10.2136/sssaj2017.08.0303

Dillon, P. (2005). Future management of aquifer recharge. *Hydrogeology Journal*, *13*(1), 313–316. https://doi.org/10.1007/s10040-004-0413-6

Dillon, P., Stuyfzand, P., Grischek, T., Lluria, M., Pyne, R. D. G., Jain, R. C., Bear, J., Schwarz, J., Wang, W., Fernandez, E., Stefan, C., Pettenati, M., van der Gun, J., Sprenger, C., Massmann, G., Scanlon, B. R., Xanke, J., Jokela, P., Zheng, Y., … Sapiano, M. (2019). Sixty years of global progress in managed aquifer recharge. *Hydrogeology Journal*, *27*(1), 1–30. https://doi.org/10.1007/s10040-018-1841-z

Elkayam, R. (2019). *Shafdan Soil Aquifer Treatment System; Process Assessment & Improvement*. the Hebrew university of Jerusalem.

Elkayam, R., Aharoni, A., Vaizel-Ohayon, D., Sued, O., Katz, Y., Negev, I., Marano, R. B. M., Cytryn, E., Shtrasler, L., & Lev, O. (2018). Viral and Microbial Pathogens, Indicator Microorganisms, Microbial Source Tracking Indicators, and Antibiotic Resistance Genes in a Confined Managed Effluent Recharge System. *Journal of Environmental Engineering (United States)*, *144*(3). https://doi.org/10.1061/(ASCE)EE.1943-7870.0001334

Elkayam, R., Michail, M., Mienis, O., Kraitzer, T., Tal, N., & Lev, O. (2015). Soil Aquifer Treatment as Disinfection Unit. *Journal of Environmental Engineering (United States)*, *141*(12). https://doi.org/10.1061/(ASCE)EE.1943-7870.0000992

Elkayam, R., Sopliniak, A., Gasser, G., Pankratov, I., & Lev, O. (2015). Oxidizer demand in the unsaturated zone of a surface-spreading soil aquifer treatment system. *Vadose Zone Journal*, *14*(11). https://doi.org/10.2136/vzj2015.03.0047

Feng, G. L., Letey, J., & Wu, L. (2001). Water Ponding Depths Affect Temporal Infiltration Rates in a Water-Repellent Sand. *Soil Science Society of America Journal*, *65*(2), 315–320. https://doi.org/10.2136/sssaj2001.652315x

Furman, A., Warrick, A. W., Zerihun, D., & Sanchez, C. A. (2006). Modified Kostiakov Infiltration Function: Accounting for Initial and Boundary Conditions. *Journal of Irrigation and Drainage Engineering*, *132*(6), 587–596. https://doi.org/10.1061/(ASCE)0733-9437(2006)132:6(587)

Girei, A., Nabayi, A., Aliyu, J., Garba, J., Hashim, S., Alasinri, S., & Abdullahi, M. (2019). Performance of Horton Infiltration model in Predicting the Infiltration Capacity of some Soils of the Sudan Savanna of Nigeria. *Nigerian Journal of Soil Science*, 10–16. https://doi.org/10.36265/njss.2019.290102

Goren, O., Burg, A., Gavrieli, I., Negev, I., Guttman, J., Kraitzer, T., Kloppmann, W., & Lazar, B. (2014). Biogeochemical processes in infiltration basins and their impact on the recharging effluent, the soil aquifer treatment (SAT) system of the Shafdan plant, Israel. *Applied Geochemistry*, *48*, 58–69. https://doi.org/10.1016/j.apgeochem.2014.06.017

Grant, S. B., Saphores, J.-D., Feldman, D. L., Hamilton, A. J., Fletcher, T. D., Cook, P. L. M., Stewardson, M., Sanders, B. F., Levin, L. A., Ambrose, R. F., Deletic, A., Brown, R., Jiang, S. C., Rosso, D., Cooper, W. J., & Marusic, I. (2012). Taking the "Waste" Out of "Wastewater" for Human Water Security and Ecosystem Sustainability. *Science*, *337*(6095), 681–686. https://doi.org/10.1126/science.1216852

Green, W. H., & Ampt, G. (1911). Studies on Soil Phyics. *The Journal of Agricultural Science*, *4*(1), 1–24.

Grinshpan, M., Furman, A., Dahlke, H. E., Raveh, E., & Weisbrod, N. (2021). From managed aquifer recharge to soil aquifer treatment on agricultural soils: Concepts and challenges. *Agricultural Water Management*, *255*, 106991. https://doi.org/10.1016/j.agwat.2021.106991

Haaken, K., Furman, A., Weisbrod, N., & Kemna, A. (2016). Time-Lapse Electrical Imaging of Water Infiltration in the Context of Soil Aquifer Treatment. *Vadose Zone Journal*, *15*(11), 1–12. https://doi.org/10.2136/vzj2016.04.0028

Hochman, A., Kunin, P., Alpert, P., Harpaz, T., Saaroni, H., & Rostkier-Edelstein, D. (2020). Weather regimes and analogues downscaling of seasonal precipitation for the 21st century: A case study over Israel. *International Journal of Climatology*, *40*(4), 2062–2077. https://doi.org/10.1002/joc.6318

Horton. R. E. (1940). The infiltration-theory of surface-runoff. *Eos, Transactions American Geophysical Union*, *21*(2), 541–541. https://doi.org/10.1029/TR021i002p00541-1

Hossain-Anni, A., Cohen, S., & Praskievicz, S. (2020). Sensitivity of urban flood simulations to stormwater infrastructure and soil infiltration. *Journal of Hydrology*, *588*, 125028. https://doi.org/10.1016/j.jhydrol.2020.125028

Katznelson, R. (1989). Clogging of groundwater recharge basins by cyanobacterial mats. *FEMS Microbiology Letters*, *62*(4), 231–242. https://doi.org/10.1016/0378-1097(89)90247-4

Kostiakov A.N. (1932). *On the dynamics of the coefficient of water percolation in soils and on the necessity for studying it from a dynamic point of view for purposes of amelioration*. *6*(17–21).

Kümmerer, K., Dionysiou, D. D., Olsson, O., & Fatta-Kassinos, D. (2018). A path to clean water. *Science*, *361*(6399), 222–224. https://doi.org/10.1126/science.aau2405

Lake, H. R., Akbarzadeh, A., & Mehrjardi, R. T. (2009). Development of pedo transfer functions (PTFs) to predict soil physico-chemical and hydrological characteristics in southern coastal zones of the Caspian Sea. *Journal of Ecology and the Natural Environment*, *1*(7), 160–172.

Lee, B.-J., Lee, J.-H., Yoon, H., & Lee, E. (2015). Hydraulic Experiments for Determination of In-situ Hydraulic Conductivity of Submerged Sediments. *Scientific Reports*, *5*(1), 7917. https://doi.org/10.1038/srep07917

Li, Z., Chen, M., Gao, S., Wen, Y., Gourley, J. J., Yang, T., Kolar, R., & Hong, Y. (2022). Can re-infiltration process be ignored for flood inundation mapping and prediction during extreme storms? A case study in Texas Gulf Coast region. *Environmental Modelling & Software*, *155*, 105450. https://doi.org/10.1016/j.envsoft.2022.105450

Magesan, G. N., Dalgety, J., Lee, R., Luo, J., & van Oostrom, A. J. (1999). Preferential Flow and Water Quality in Two New Zealand Soils Previously Irrigated with Wastewater. *Journal of Environmental Quality*, *28*(5), 1528–1532. https://doi.org/10.2134/jeq1999.00472425002800050018x

Mahapatra, S., Jha, M. K., Biswal, S., & Senapati, D. (2020). Assessing Variability of Infiltration Characteristics and Reliability of Infiltration Models in a Tropical Sub-humid Region of India. *Scientific Reports*, *10*(1), 1515. https://doi.org/10.1038/s41598-020-58333-8

Maliva, R. G. (2020). *Anthropogenic Aquifer Recharge and Water Quality* (pp. 133–164). https://doi.org/10.1007/978-3-030-11084-0_6

Mansell, J., & Drewes, J. E. (2004). Fate of Steroidal Hormones During Soil-aquifer Treatment. *Gorund Water Monitoring e Remediation*, *24*(2), 94–101. https://doi.org/10.1111/j.1745-6592.2004.tb00717.x

Mienis, O. (2013). *The influence of operative parameters on infiltration in infiltration field of secondary effluent (In Hebrew)*. The Hebrew University of Jerusalem.

Mizrahi, G., Furman, A., & Weisbrod, N. (2016). Infiltration under Confined Air Conditions: Impact of Inclined Soil Surface. *Vadose Zone Journal*, *15*(9), 1–8. https://doi.org/10.2136/vzj2016.04.0034

Nadav, I., Tarchitzky, J., & Chen, Y. (2012). Soil cultivation for enhanced wastewater infiltration in soil aquifer treatment (SAT). *Journal of Hydrology*, *470–471*, 75–81. https://doi.org/10.1016/j.jhydrol.2012.08.013

Page, D., Bekele, E., Vanderzalm, J., & Sidhu, J. (2018). Managed Aquifer Recharge (MAR) in Sustainable Urban Water Management. *Water*, *10*(3), 239. https://doi.org/10.3390/w10030239

Philip, J. R. (1969). *Theory of Infiltration* (pp. 215–296). https://doi.org/10.1016/B978-1-4831-9936-8.50010-6

Raats, P. A. C., & Knight, J. H. (2018). The Contributions of Lewis Fry Richardson to Drainage Theory, Soil Physics, and the Soil-Plant-Atmosphere Continuum. *Frontiers in Environmental Science*, *6*. https://doi.org/10.3389/fenvs.2018.00013

Racz, A. J., Fisher, A. T., Schmidt, C. M., Lockwood, B. S., & Huertos, M. L. (2012). Spatial and Temporal Infiltration Dynamics During Managed Aquifer Recharge. *Groundwater*, *50*(4), 562–570. https://doi.org/10.1111/j.1745-6584.2011.00875.x

Rice, R. C., & Bouwer, H. (2013). Soil aquifer treatment using primary effluent. *Water Environment Federation*.

Rice, R. C., & Rice, C. (2013). *Soil Clogging during Infiltration of Secondary Effluent*. *46*(4), 708–716.

Richards, L. A. (1931). Capillary Conduction of Liquids Through Porous Mediums. *Physics*, *1*(5), 318–333. https://doi.org/10.1063/1.1745010

Schwabe, K., Nemati, M., Amin, R., Tran, Q., & Jassby, D. (2020). Unintended consequences of water conservation on the use of treated municipal wastewater. *Nature Sustainability*, *3*(8), 628–635. https://doi.org/10.1038/s41893-020-0529-2

Shapiro, S. S., & Wilk, M. B. (1965). An analysis of variance test for normality (complete samples). *Biometrika*, *52*(3–4), 591–611. https://doi.org/10.1093/biomet/52.3-4.591

Shukla, M. K., Lal, R., & Unkefer, P. (2003). Experimental Evaluation of Infiltration Models for Various Land Use and Soil Management Systems. *Soil Science*, *168*(3), 178–191. https://doi.org/10.1097/01.ss.0000058890.60072.7c

Sihag, P. (2018). Prediction of unsaturated hydraulic conductivity using fuzzy logic and artificial neural network. *Modeling Earth Systems and Environment*, *4*(1), 189--198.

Sihag, P., Tiwari, N. K., & Ranjan, S. (2017). Estimation and inter-comparison of infiltration models. *Water Science*, *31*(1), 34–43. https://doi.org/10.1016/j.wsj.2017.03.001

Talsma, T., & Parlange, J. (1972). One dimensional vertical infiltration. *Soil Research*, *10*(2), 143. https://doi.org/10.1071/SR9720143

Tokunaga, T. K. (2020). Simplified Green-Ampt Model, Imbibition-Based Estimates of Permeability, and Implications for Leak-off in Hydraulic Fracturing. *Water Resources Research*, *56*(4). https://doi.org/10.1029/2019WR026919

Vinutha, H. P., Poornima, B., & Sagar, B. M. (2018). *Detection of Outliers Using Interquartile Range Technique from Intrusion Dataset* (pp. 511–518). https://doi.org/10.1007/978-981-10-7563-6_53

Warrick, A. W., Zerihun, D., Sanchez, C. A., & Furman, A. (2005). Infiltration under Variable Ponding Depths of Water. *Journal of Irrigation and Drainage Engineering*, *131*(4), 358–363. https://doi.org/10.1061/(ASCE)0733-9437(2005)131:4(358)

Ying, G.-G., Kookana, R. S., & Dillon, P. (2003). Sorption and degradation of selected five endocrine disrupting chemicals in aquifer material. *Water Research*, *37*(15), 3785–3791. https://doi.org/10.1016/S0043-1354(03)00261-6

Zheng, Y., Ross, A., Villholth, K. G., & Dillon, P. (2021). *Managing aquifer recharge: a showcase for resilience and sustainability*. UNESCO, IAH, and GRIPP.

SUPPLEMENTARY INFORMATION

# Head-Independent Time-Invariant Infiltration Rate in Aquifer Recharge with Treated Municipal Wastewater

Roy Elkayam[*,†,‡] and Ovadia Lev[‡]

†Mekorot Water Company, Lincoln Street, Tel-Aviv – Yafo, 6492105, Israel.

The Institute of Chemistry, The Hebrew University of Jerusalem, Jerusalem, 9190401, Israel

E-mail: relkayam@mekorot.co.il

# Content

S1: Overview of the Israeli Coastal Aquifer and Shafdan SAT System:

## Figures

Figure S1: Lithology of observation wells drilled in several lagoons in the Shafdan SAT.
Figure S2: Examples of Outliers in Basin Operation Data.
*Figure S3: Flow rate fluctuations to the Shafdan SAT basins during typical flooding sessions.*
*Figure S4: Typical distribution of the relative standard deviations.*
*Figure S5: Histograms of the distributions of the slope and coefficient of determination of the relationship*

## Tables

Table S1: Statistical attributes of the linear dependence of $Ir_d$ on H during the drainage phases in all 50 studied lagoons for the last decade

Table S2: Linear correlations between the average infiltration rate during the wetting phases $\overline{Ir_w}$ and the head-independent infiltration rates during the drainage phase, $Ir_d$ in all 50 studied basins for the last decade

# S1: Overview of the Israeli Coastal Aquifer and Shafdan SAT System:

The Israeli coastal aquifer, which stretches along the Mediterranean Sea in the western part of the country, is composed of Pleistocene-age rocks from a coastal environment, mainly consisting of calcareous sandstone (the Kurkar group). This aquifer developed through a series of depositional cycles and is distinguished by the alternation of sandstone layers with lenses of marine and continental clay, silt, and shale. These represent the fluctuating sea level stages during the glacial and interglacial periods of the Pleistocene epoch. [1,2] The sandstone formation of the Kurkar Group continues to accumulate on the shallow continental shelf today [3].

The coastal aquifer overlies the late Eocene to early Pleistocene Saqiye Group, a thick, impermeable clayish unit sloping westward. As a result, the overlying aquifer attains its maximum thickness (180–200 m) along the coastline and gradually thins out eastward, disappearing about 10–15 km inland. The vadose zone beneath the infiltration ponds is approximately 30–40 m thick.

Three north–south trending, sub-parallel calcareous sandstone (kurkar) ridges, located roughly 0.5–0.7, 1.3–2, and 3–3.5 km east of the coastline and separated by clayey elongated basins, dominate the aquifer's morphology. The recharge ponds are predominantly constructed on the Eastern Ridge. In certain areas, thicker marine clay layers divide the aquifer into sub-horizontal sections. This division is more pronounced toward the coast and affects the western parts of the aquifer. Despite its irregularity, four main sub-aquifers, situated on top of each other, can be identified from top to bottom as A–D.

The Shafdan SAT system predominantly draws water from sub-aquifer B, which is typically 60–80 m thick and characterized mainly by sandstone and conglomerate with relatively high hydraulic conductivities. Typically, the conductivity is in the range of 4–10 m/s[4].

The recharge wells (RWs) are arranged in a double-ring structure around the percolation ponds. The inner ring pumps 100% treated effluent, while the outer ring pumps 70–85% of treated effluent along with a complementary amount of natural, regional aquifer water. The number, locations, capacities, screen positions, and other features of the wells were designed to establish a closed Subsurface-Aeration Trenches (SAT) system, which is isolated from the surrounding freshwater aquifer and does not impact its chemical quality.

The first infiltration basins (Soreq) were established west of Rishon Lezion in 1977.[5] They were designed to treat 50 Mm3/year of effluent. The SAT system has expanded continuously to meet demand, with additional facilities constructed west of Yavne and north of Ashdod (Fig. 2 in the main article). Currently, the overall recharge of secondary effluent is approximately 130 Mm3/year, while reclaimed water totals around 145 Mm3/year and is pumped by about 150 recovery wells[6].

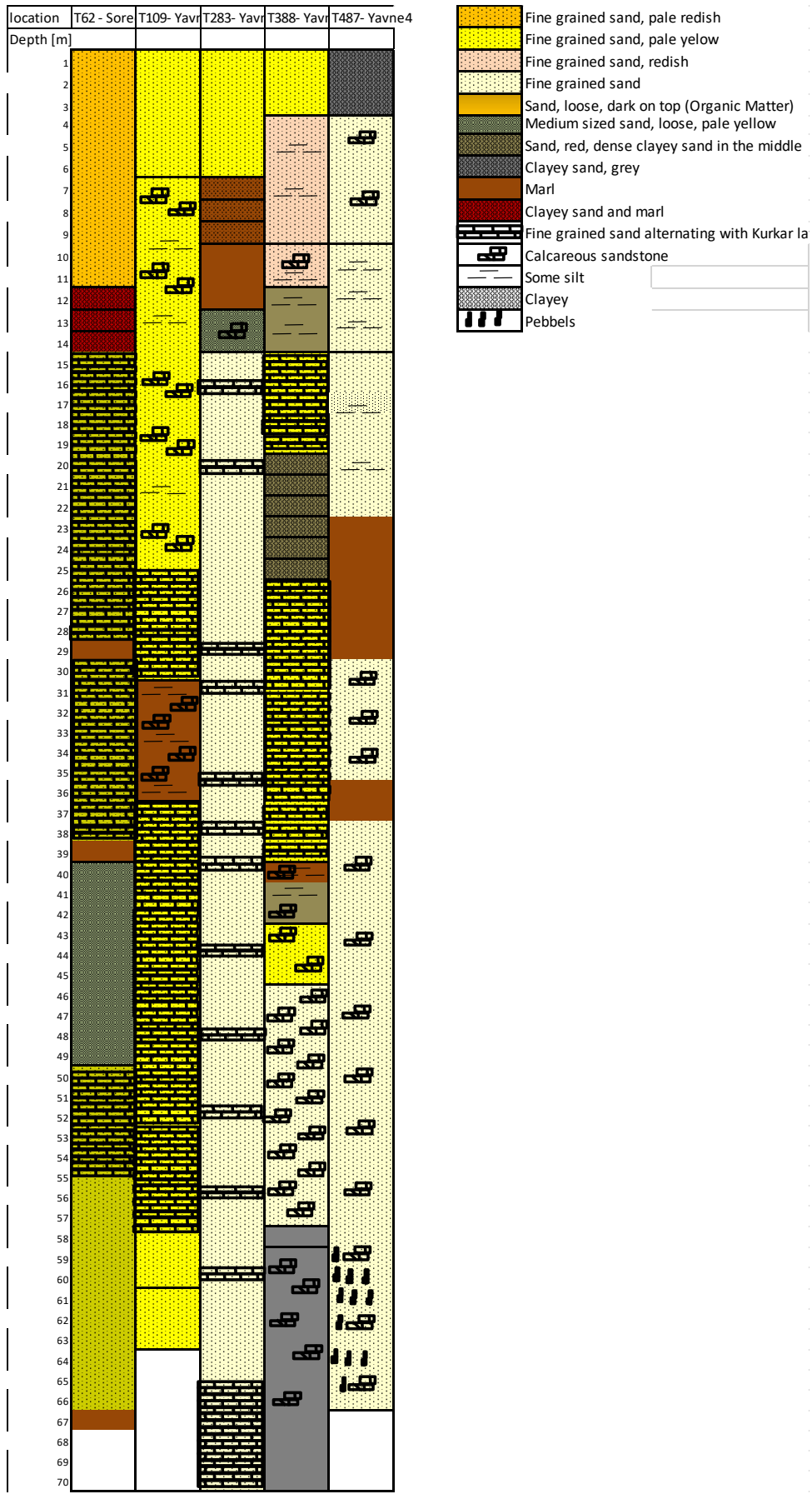

*Figure S1: Lithology of observation wells drilled in several lagoons in the Shafdan SAT.*

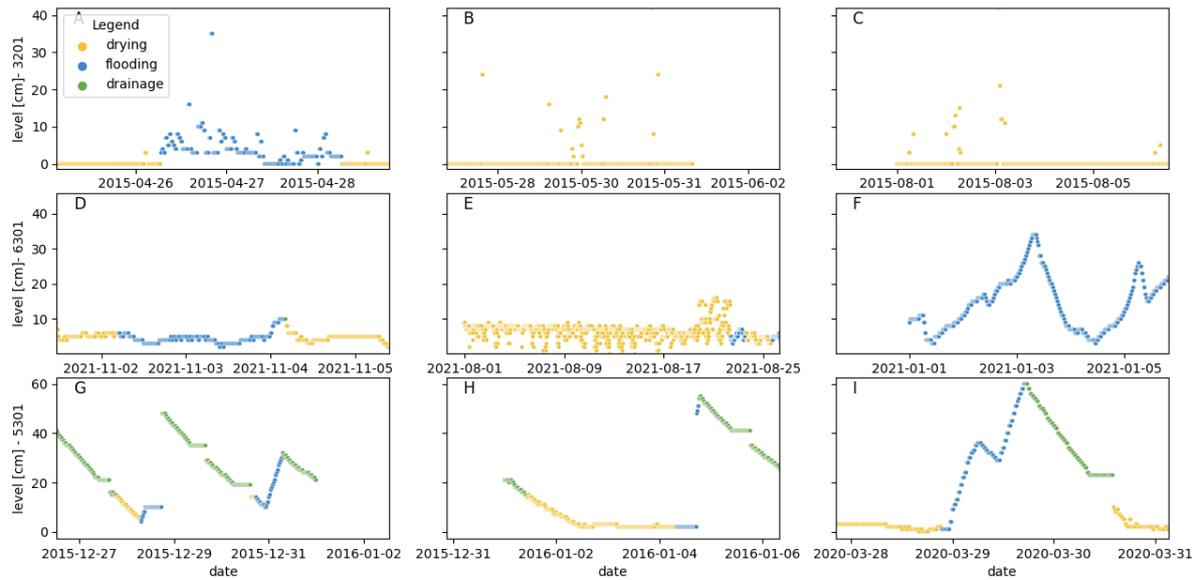

*Figure S2: Examples of Outliers in Basin Operation Data. A) A 24-hour flooding period that does not sum up to a proper flooding event, possibly due to an insignificant change in basin water level or a fault in the level measuring instrument. B+C) Instances indicating a potential fault in the level measuring instrument. D+E) Insignificant flooding events or potential faults in the level measuring instrument. F) A valve status fault showing the basin as open when the data clearly indicates a flooding event, with identifiable flooding and drainage statuses. G+H+I) Cases where communication faults result in frozen values in the basin level measurements.*

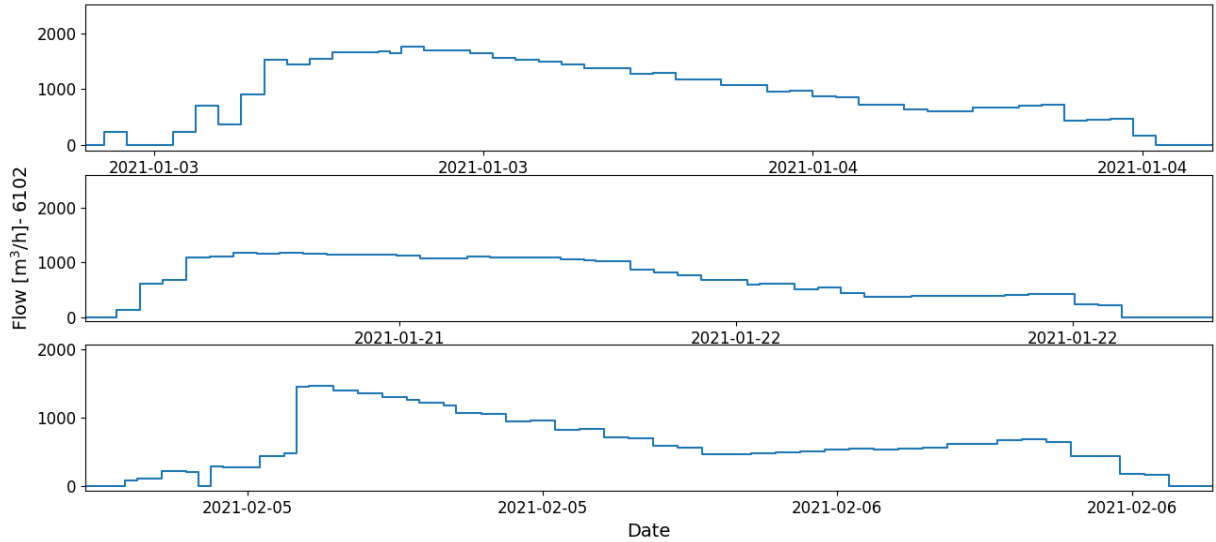

Figure S3: Flow rate fluctuations to the Shafdan SAT basins during typical flooding sessions. The fluctuations are not necessarily intentional, they can be caused by opening and closing of inlet valves to other basins.

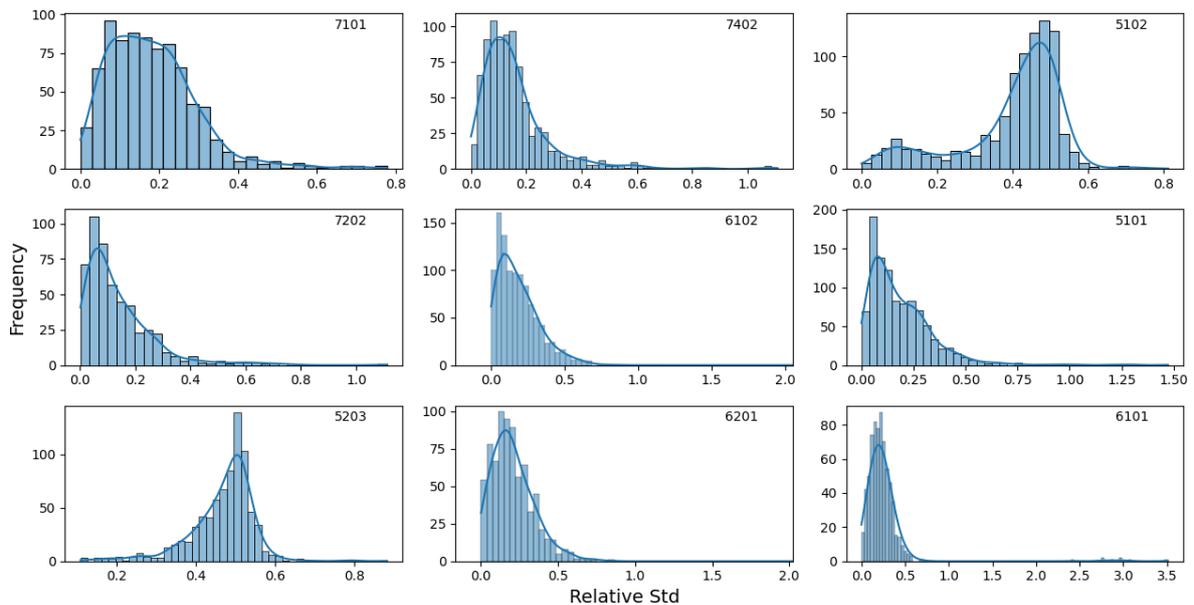

Figure S4: Typical distribution of the relative standard deviations (normalized by the average flow rate during each cycle) of the flow rates ($m^3/m^2$) to the respective basins in all the proper cycles during the studied decade.

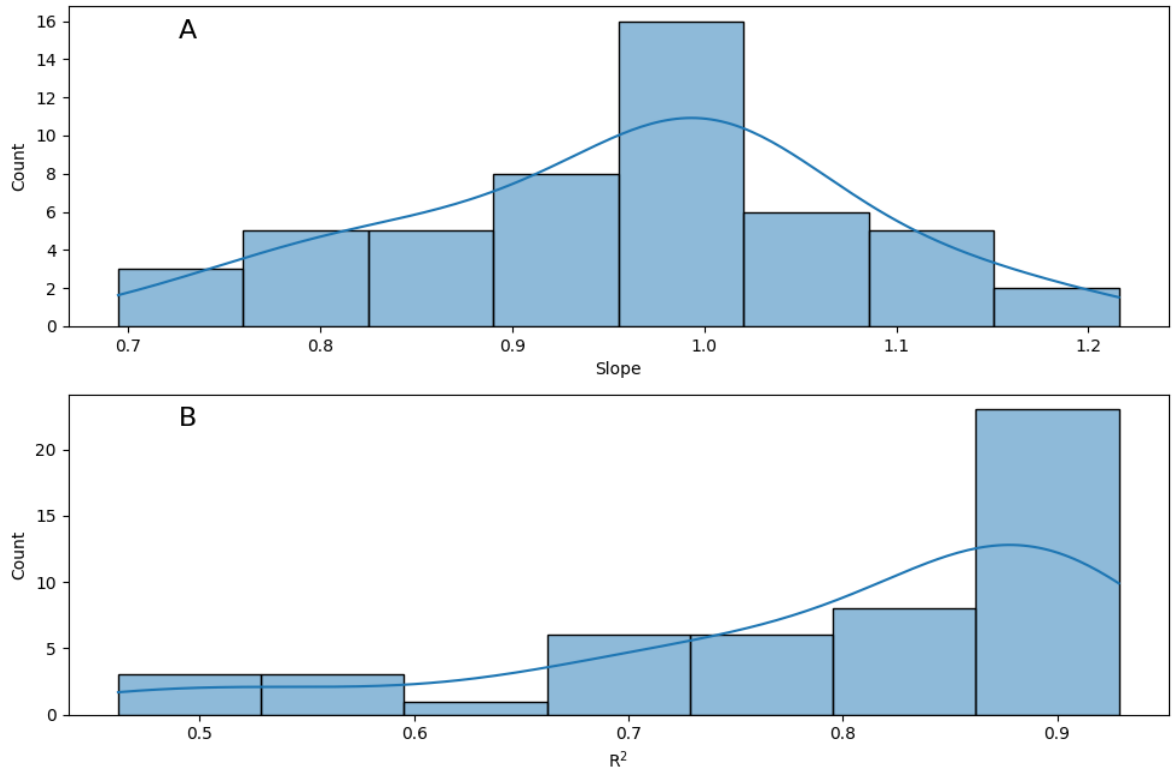

*Figure S5: Histograms of the distributions of the slope and coefficient of determination of the relationship ($\overline{r_w}$ = k $Ir_d$). Upper frame, distribution of observed slope, k. Lower frame distribution of the coefficients of determination, $R^2$.*

**Table S1:** Statistical attributes of the linear dependence of $Ir_d$ on H during the drainage phases in all 50 studied lagoons for the last decade.

| Basin | N | Coefficient Of Determination, $R^2$ | | | | $C_2$, Pre-logarithmic constant in equation 8 | | | | p-value for average $C_2<0$ | |
|---|---|---|---|---|---|---|---|---|---|---|---|
| | | Max | Min | Avg | Std | Max | Min | Avg | Std | p-value, Avg | p-value, Std |
| 3201 | 701 | 0.998 | 0.940 | 0.982 | 0.016 | -0.004 | -0.038 | -0.015 | 0.005 | 3E-06 | 3E-05 |
| 3202 | 955 | 0.998 | 0.940 | 0.989 | 0.011 | -0.005 | -0.050 | -0.019 | 0.006 | 4E-06 | 4E-05 |
| 3203 | 1131 | 0.998 | 0.941 | 0.993 | 0.005 | -0.004 | -0.042 | -0.018 | 0.006 | 3E-07 | 7E-06 |
| 3204 | 1090 | 0.998 | 0.941 | 0.985 | 0.009 | -0.006 | -0.035 | -0.017 | 0.005 | 2E-06 | 2E-05 |
| 4101 | 751 | 0.999 | 0.940 | 0.993 | 0.007 | -0.001 | -0.024 | -0.007 | 0.003 | 2E-10 | 5E-09 |
| 4102 | 830 | 0.998 | 0.940 | 0.989 | 0.010 | -0.001 | -0.030 | -0.009 | 0.004 | 3E-06 | 3E-05 |
| 4103 | 619 | 0.998 | 0.940 | 0.984 | 0.014 | -0.002 | -0.026 | -0.010 | 0.004 | 1E-06 | 1E-05 |
| 4104 | 653 | 0.999 | 0.942 | 0.992 | 0.010 | -0.002 | -0.011 | -0.005 | 0.001 | 5E-08 | 1E-06 |
| 4201 | 827 | 0.999 | 0.943 | 0.991 | 0.011 | -0.0004 | -0.021 | -0.008 | 0.004 | 2E-11 | 5E-10 |
| 4202 | 838 | 0.998 | 0.946 | 0.990 | 0.007 | -0.002 | -0.014 | -0.006 | 0.002 | 4E-08 | 1E-06 |
| 4203 | 691 | 0.999 | 0.953 | 0.992 | 0.005 | -0.001 | -0.025 | -0.007 | 0.004 | 8E-08 | 1E-06 |
| 4204 | 803 | 0.999 | 0.954 | 0.992 | 0.005 | -0.002 | -0.026 | -0.008 | 0.004 | 6E-13 | 8E-12 |
| 4301 | 918 | 0.998 | 0.942 | 0.991 | 0.008 | -0.003 | -0.030 | -0.011 | 0.005 | 3E-09 | 9E-08 |
| 4302 | 892 | 0.998 | 0.941 | 0.984 | 0.013 | -0.002 | -0.023 | -0.012 | 0.004 | 3E-09 | 8E-08 |
| 4303 | 501 | 0.999 | 0.943 | 0.989 | 0.007 | -0.001 | -0.013 | -0.003 | 0.001 | 4E-09 | 1E-07 |
| 4304 | 566 | 0.998 | 0.943 | 0.990 | 0.008 | -0.001 | -0.008 | -0.003 | 0.001 | 1E-16 | 2E-15 |
| 4401 | 882 | 0.999 | 0.941 | 0.992 | 0.008 | -0.002 | -0.030 | -0.010 | 0.005 | 2E-08 | 6E-07 |
| 4402 | 777 | 0.999 | 0.940 | 0.985 | 0.013 | -0.001 | -0.032 | -0.011 | 0.004 | 8E-08 | 2E-06 |
| 4403 | 807 | 0.999 | 0.942 | 0.995 | 0.006 | -0.001 | -0.023 | -0.006 | 0.003 | 4E-07 | 9E-06 |
| 4404 | 637 | 0.999 | 0.950 | 0.993 | 0.005 | -0.001 | -0.027 | -0.004 | 0.002 | 1E-12 | 3E-11 |
| 5101 | 1026 | 0.998 | 0.947 | 0.993 | 0.005 | -0.003 | -0.028 | -0.012 | 0.004 | 5E-08 | 1E-06 |
| 5102 | 903 | 0.999 | 0.945 | 0.994 | 0.005 | -0.004 | -0.030 | -0.012 | 0.004 | 2E-07 | 4E-06 |
| 5103 | 746 | 0.999 | 0.941 | 0.995 | 0.007 | -0.003 | -0.034 | -0.011 | 0.005 | 5E-07 | 1E-05 |
| 5201 | 955 | 0.998 | 0.940 | 0.989 | 0.010 | -0.003 | -0.043 | -0.014 | 0.006 | 7E-07 | 1E-05 |
| 5202 | 990 | 0.996 | 0.941 | 0.985 | 0.009 | -0.006 | -0.044 | -0.020 | 0.007 | 1E-06 | 1E-05 |
| 5203 | 781 | 0.998 | 0.940 | 0.990 | 0.010 | -0.003 | -0.034 | -0.011 | 0.004 | 7E-07 | 1E-05 |
| 5301 | 725 | 0.999 | 0.942 | 0.993 | 0.008 | -0.001 | -0.017 | -0.006 | 0.002 | 4E-07 | 1E-05 |
| 5302 | 686 | 0.999 | 0.941 | 0.993 | 0.010 | -0.002 | -0.037 | -0.007 | 0.003 | 9E-08 | 2E-06 |
| 5303 | 949 | 0.998 | 0.940 | 0.990 | 0.007 | -0.006 | -0.030 | -0.016 | 0.004 | 6E-07 | 9E-06 |
| 6101 | 696 | 0.998 | 0.941 | 0.988 | 0.013 | -0.005 | -0.033 | -0.016 | 0.006 | 3E-06 | 2E-05 |
| 6102 | 1035 | 0.999 | 0.943 | 0.995 | 0.005 | -0.004 | -0.036 | -0.014 | 0.006 | 1E-07 | 3E-06 |
| 6103 | 819 | 0.999 | 0.946 | 0.993 | 0.006 | -0.003 | -0.040 | -0.012 | 0.005 | 4E-07 | 1E-05 |
| 6201 | 760 | 0.999 | 0.941 | 0.993 | 0.007 | -0.002 | -0.038 | -0.012 | 0.006 | 9E-07 | 2E-05 |
| 6202 | 757 | 0.999 | 0.940 | 0.991 | 0.010 | -0.003 | -0.033 | -0.013 | 0.006 | 2E-06 | 2E-05 |
| 6203 | 608 | 0.998 | 0.942 | 0.988 | 0.011 | -0.004 | -0.060 | -0.014 | 0.007 | 2E-06 | 2E-05 |
| 6301 | 521 | 0.998 | 0.941 | 0.991 | 0.011 | -0.003 | -0.035 | -0.013 | 0.006 | 3E-06 | 2E-05 |
| 6302 | 715 | 0.999 | 0.941 | 0.994 | 0.007 | -0.004 | -0.032 | -0.011 | 0.005 | 2E-07 | 3E-06 |
| 6303 | 643 | 0.999 | 0.941 | 0.992 | 0.008 | -0.003 | -0.025 | -0.010 | 0.005 | 2E-06 | 2E-05 |
| 7101 | 810 | 0.999 | 0.941 | 0.994 | 0.008 | -0.004 | -0.034 | -0.012 | 0.005 | 1E-06 | 2E-05 |
| 7102 | 251 | 0.999 | 0.947 | 0.991 | 0.010 | -0.001 | -0.005 | -0.002 | 0.001 | 1E-19 | 2E-18 |
| 7103 | 519 | 0.999 | 0.940 | 0.989 | 0.011 | -0.001 | -0.013 | -0.003 | 0.001 | 5E-07 | 1E-05 |
| 7201 | 449 | 0.999 | 0.944 | 0.995 | 0.006 | -0.001 | -0.007 | -0.003 | 0.001 | 2E-26 | 4E-25 |
| 7202 | 518 | 0.997 | 0.941 | 0.987 | 0.011 | -0.001 | -0.006 | -0.002 | 0.001 | 2E-15 | 5E-14 |
| 7203 | 367 | 0.998 | 0.943 | 0.988 | 0.010 | -0.001 | -0.009 | -0.003 | 0.001 | 1E-14 | 3E-13 |
| 7301 | 604 | 0.999 | 0.946 | 0.993 | 0.006 | -0.001 | -0.010 | -0.005 | 0.002 | 5E-13 | 1E-11 |
| 7302 | 648 | 0.998 | 0.941 | 0.981 | 0.015 | -0.003 | -0.034 | -0.012 | 0.006 | 2E-06 | 2E-05 |
| 7303 | 642 | 0.999 | 0.941 | 0.991 | 0.011 | -0.001 | -0.032 | -0.005 | 0.003 | 6E-07 | 1E-05 |
| 7401 | 729 | 0.999 | 0.944 | 0.993 | 0.009 | -0.001 | -0.014 | -0.005 | 0.002 | 2E-09 | 5E-08 |
| 7402 | 854 | 0.999 | 0.941 | 0.990 | 0.012 | -0.004 | -0.037 | -0.014 | 0.006 | 3E-06 | 3E-05 |
| 7403 | 801 | 0.998 | 0.941 | 0.984 | 0.013 | -0.004 | -0.039 | -0.012 | 0.005 | 3E-07 | 4E-06 |

*The columns depict the maximal, minimal, average COD, $R^2$ and its standards deviation, obtained in the logarithmic fit of log ($Ir_d$) vs. log (H) (Equation 8) for each of 50 examined basins. The statistical characteristics of $C_2$ are also depicted with the maximal, minimal, average and standard deviation of the slope of the logarithmic dependencies in each basin. p-value avg. represents p-values of rejecting the null hypothesis of average $C_2 \geq 0$ for each basin. p-value, Std values represent the standard deviations of p-value in all the studied cycles of each basin.*
.

Table S2: Linear correlations between the average infiltration rate during the wetting phases ($\overline{Ir_w}$) and the head-independent infiltration rates during the drainage phase, $Ir_d$ in all 50 studied basins for the last decade.

| Field | Basin | N Events | Outliers | | Linear correlation (Irw = slope·Ird + Y-intercept) | | | | | Constrained fit (Irw = slope·Ird) | | | Shapiro-Wilk Normality test | |
|---|---|---|---|---|---|---|---|---|---|---|---|---|---|---|
| | | | Event | Removal % | slope | Y-intercept | std error | srMSE | $R^2$ | slope | srMSE | $R^2$ | Statistic | p-value |
| Soreq | 3201 | 1068 | 63 | 6% | 0.590 | 1.047 | 0.014 | 0.138 | 0.639 | 0.843 | 0.160 | 0.513 | 0.725 | 1.11E-38 |
| | 3202 | 1176 | 18 | 2% | 0.838 | 0.689 | 0.012 | 0.115 | 0.811 | 0.961 | 0.121 | 0.793 | 0.306 | 0.00E+00 |
| | 3203 | 1270 | 7 | 1% | 0.979 | 0.192 | 0.008 | 0.087 | 0.925 | 1.011 | 0.088 | 0.923 | 0.807 | 2.11E-36 |
| | 3204 | 1263 | 12 | 1% | 0.999 | 0.727 | 0.012 | 0.093 | 0.851 | 1.185 | 0.103 | 0.819 | 0.809 | 3.31E-36 |
| Yavne 1 | 4101 | 891 | 60 | 7% | 1.052 | 0.025 | 0.013 | 0.136 | 0.892 | 1.061 | 0.136 | 0.891 | 0.717 | 2.88E-36 |
| | 4102 | 1051 | 92 | 9% | 1.172 | 0.017 | 0.017 | 0.183 | 0.826 | 1.178 | 0.183 | 0.826 | 0.841 | 3.99E-31 |
| | 4103 | 890 | 101 | 11% | 0.693 | 0.559 | 0.015 | 0.168 | 0.733 | 0.853 | 0.182 | 0.689 | 0.927 | 2.53E-20 |
| | 4104 | 781 | 30 | 4% | 0.956 | 0.242 | 0.023 | 0.178 | 0.692 | 1.104 | 0.183 | 0.674 | 0.906 | 1.49E-21 |
| | 4201 | 957 | 71 | 7% | 1.099 | -0.112 | 0.010 | 0.110 | 0.933 | 1.061 | 0.111 | 0.932 | 0.660 | 6.18E-40 |
| | 4202 | 963 | 136 | 14% | 1.231 | -0.376 | 0.020 | 0.178 | 0.818 | 1.083 | 0.184 | 0.805 | 0.759 | 2.74E-35 |
| | 4203 | 784 | 39 | 5% | 1.008 | 0.016 | 0.012 | 0.140 | 0.900 | 1.013 | 0.140 | 0.900 | 0.542 | 6.28E-41 |
| | 4204 | 900 | 24 | 3% | 1.308 | -0.284 | 0.014 | 0.135 | 0.909 | 1.217 | 0.139 | 0.904 | 0.065 | 0.00E+00 |
| | 4301 | 1221 | 176 | 14% | 0.990 | 0.107 | 0.016 | 0.188 | 0.795 | 1.014 | 0.188 | 0.794 | 0.355 | 0.00E+00 |
| | 4302 | 1181 | 120 | 10% | 1.091 | -0.011 | 0.014 | 0.159 | 0.853 | 1.088 | 0.159 | 0.853 | 0.189 | 0.00E+00 |
| | 4303 | 566 | 51 | 9% | 1.012 | -0.060 | 0.023 | 0.199 | 0.786 | 0.969 | 0.199 | 0.785 | 0.056 | 0.00E+00 |
| | 4304 | 648 | 103 | 16% | 0.817 | 0.081 | 0.028 | 0.201 | 0.603 | 0.893 | 0.202 | 0.597 | 0.970 | 3.08E-10 |
| | 4401 | 1034 | 85 | 8% | 0.990 | -0.039 | 0.014 | 0.218 | 0.845 | 0.980 | 0.218 | 0.845 | 0.064 | 0.00E+00 |
| | 4402 | 1029 | 114 | 11% | 0.812 | 0.350 | 0.021 | 0.249 | 0.618 | 0.904 | 0.252 | 0.609 | 0.600 | 1.77E-43 |
| | 4403 | 913 | 202 | 22% | 0.775 | 0.088 | 0.022 | 0.260 | 0.633 | 0.808 | 0.260 | 0.632 | 0.126 | 0.00E+00 |
| | 4404 | 747 | 98 | 13% | 0.900 | 0.018 | 0.022 | 0.271 | 0.718 | 0.912 | 0.271 | 0.718 | 0.639 | 6.92E-37 |
| Yavne 2 | 5101 | 1169 | 52 | 4% | 1.081 | -0.275 | 0.010 | 0.107 | 0.918 | 0.989 | 0.112 | 0.910 | 0.793 | 4.27E-36 |
| | 5102 | 1015 | 32 | 3% | 1.087 | -0.072 | 0.010 | 0.097 | 0.925 | 1.067 | 0.097 | 0.924 | 0.355 | 0.00E+00 |
| | 5103 | 856 | 23 | 3% | 1.078 | -0.224 | 0.012 | 0.136 | 0.901 | 1.015 | 0.139 | 0.897 | 0.866 | 2.24E-26 |

| Field | Basin | N Event | Outliers | % | Slope | Constant | Std Err | srMSE | R² | Slope (c) | srMSE (c) | R² (c) | Shapiro-Wilk | p-value |
|---|---|---|---|---|---|---|---|---|---|---|---|---|---|---|
| | 5201 | 1209 | 85 | 7% | 1.001 | -0.024 | 0.011 | 0.166 | 0.877 | 0.996 | 0.166 | 0.876 | 0.272 | 0.00E+00 |
| | 5202 | 1195 | 214 | 18% | 0.934 | -0.084 | 0.013 | 0.157 | 0.850 | 0.918 | 0.157 | 0.850 | 0.914 | 1.79E-25 |
| | 5203 | 976 | 48 | 5% | 0.963 | 0.225 | 0.013 | 0.169 | 0.855 | 1.024 | 0.172 | 0.851 | 0.773 | 1.10E-34 |
| | 5301 | 824 | 104 | 13% | 1.080 | -0.112 | 0.015 | 0.131 | 0.878 | 1.028 | 0.132 | 0.876 | 0.104 | 0.00E+00 |
| | 5302 | 887 | 120 | 14% | 0.972 | 0.077 | 0.017 | 0.157 | 0.814 | 1.006 | 0.157 | 0.813 | 0.731 | 1.72E-35 |
| | 5303 | 1108 | 172 | 16% | 1.025 | -0.170 | 0.021 | 0.145 | 0.716 | 0.982 | 0.146 | 0.715 | 0.682 | 2.41E-41 |
| Yavne 3 | 6101 | 883 | 60 | 7% | 0.851 | 0.519 | 0.021 | 0.145 | 0.670 | 0.973 | 0.148 | 0.656 | 0.171 | 0.00E+00 |
| | 6102 | 1175 | 26 | 2% | 0.920 | 0.007 | 0.010 | 0.134 | 0.883 | 0.921 | 0.134 | 0.883 | 0.769 | 1.06E-37 |
| | 6103 | 934 | 29 | 3% | 1.016 | 0.022 | 0.011 | 0.129 | 0.898 | 1.022 | 0.129 | 0.898 | 0.836 | 6.41E-30 |
| | 6201 | 943 | 56 | 6% | 0.910 | 0.113 | 0.010 | 0.147 | 0.907 | 0.938 | 0.148 | 0.906 | 0.601 | 5.39E-42 |
| | 6202 | 877 | 29 | 3% | 0.885 | 0.118 | 0.011 | 0.122 | 0.877 | 0.916 | 0.123 | 0.876 | 0.693 | 3.89E-37 |
| | 6203 | 751 | 35 | 5% | 0.867 | 0.252 | 0.018 | 0.159 | 0.761 | 0.936 | 0.161 | 0.756 | 0.202 | 0.00E+00 |
| | 6301 | 635 | 24 | 4% | 0.982 | 0.083 | 0.018 | 0.134 | 0.838 | 1.003 | 0.134 | 0.837 | 0.379 | 7.55E-42 |
| | 6302 | 820 | 24 | 3% | 1.000 | -0.042 | 0.011 | 0.121 | 0.907 | 0.988 | 0.121 | 0.906 | 0.197 | 0.00E+00 |
| | 6303 | 741 | 23 | 3% | 1.179 | -0.082 | 0.016 | 0.117 | 0.880 | 1.150 | 0.117 | 0.879 | 0.651 | 2.67E-36 |
| Yavne 4 | 7101 | 958 | 114 | 12% | 0.807 | -0.179 | 0.012 | 0.155 | 0.841 | 0.753 | 0.157 | 0.837 | 0.631 | 4.02E-41 |
| | 7102 | 293 | 37 | 13% | 0.902 | -0.016 | 0.025 | 0.175 | 0.835 | 0.877 | 0.175 | 0.834 | 0.861 | 1.43E-15 |
| | 7103 | 611 | 104 | 17% | 0.796 | 0.012 | 0.022 | 0.179 | 0.730 | 0.814 | 0.179 | 0.730 | 0.948 | 8.95E-14 |
| | 7201 | 519 | 40 | 8% | 0.942 | -0.013 | 0.016 | 0.169 | 0.879 | 0.930 | 0.169 | 0.879 | 0.842 | 2.34E-22 |
| | 7202 | 641 | 113 | 18% | 0.840 | -0.019 | 0.017 | 0.150 | 0.815 | 0.807 | 0.151 | 0.813 | 0.919 | 5.43E-18 |
| | 7203 | 425 | 115 | 27% | 0.757 | 0.003 | 0.017 | 0.157 | 0.867 | 0.761 | 0.157 | 0.867 | 0.879 | 9.54E-18 |
| | 7301 | 690 | 60 | 9% | 0.873 | -0.080 | 0.011 | 0.126 | 0.910 | 0.801 | 0.131 | 0.903 | 0.475 | 8.49E-41 |
| | 7302 | 996 | 236 | 24% | 0.676 | 0.178 | 0.013 | 0.150 | 0.789 | 0.733 | 0.152 | 0.783 | 0.889 | 3.72E-26 |
| | 7303 | 832 | 82 | 10% | 0.983 | 0.043 | 0.018 | 0.257 | 0.797 | 1.009 | 0.258 | 0.796 | 0.903 | 1.63E-22 |
| | 7401 | 869 | 60 | 7% | 1.005 | -0.149 | 0.018 | 0.144 | 0.787 | 0.896 | 0.147 | 0.778 | 0.374 | 0.00E+00 |
| | 7402 | 1039 | 56 | 5% | 0.789 | 0.100 | 0.010 | 0.114 | 0.866 | 0.818 | 0.115 | 0.865 | 0.411 | 0.00E+00 |
| | 7403 | 1006 | 56 | 6% | 1.033 | 0.292 | 0.019 | 0.158 | 0.753 | 1.127 | 0.161 | 0.746 | 0.368 | 0.00E+00 |

The columns depict (left to right) the field and basin name/number, the original number of samples (N Event), the number of outliers removed, correlation results for the unconstrained regression (slope, constant, standard error, srMSE, and $R_2$), correlation results for the constrained regression (slope, srMSE, and R2). The Shapiro–Wilk normality test statistic and p-value (for rejecting the null hypothesis that the sample is taken from a normally distributed population) are given in the last columns.

srMSE = the average relative error by taking the square root of the mean square error, MSE divided by the mean infiltration rat

**REFERENCES**

(1) Issar, A. Geology of the Central Coastal Plain of Israel. *Israel Journal of Earth Sciences* **1968**, *17*, 16–29.

(2) Bruce, D.; Friedman, G. M.; Kaufman, A.; Yechieli, Y. Spatial Variations of Radiocarbon in the Coastal Aquifer Of. *Radiocarbon* **2001**, *43* (2), 783–791.

(3) Gvirtzman, G.; Buchbinder, B. Recent and Pleistocene Coral Reefs and Coastal Sediments of the Gulf of Elat. *Guidebook lOth International Congress Sedimentology* **1978**, 162–191.

(4) Kloppmann, W.; Chikurel, H.; Picot, G.; Guttman, J.; Pettenati, M.; Aharoni, A.; Guerrot, C.; Millot, R.; Gaus, I.; Wintgens, T. B and Li Isotopes as Intrinsic Tracers for Injection Tests in Aquifer Storage and Recovery Systems. *Applied Geochemistry* **2009**, *24* (7), 1214–1223. https://doi.org/10.1016/j.apgeochem.2009.03.006.

(5) Icekson-Tal, N. *Groundwater Recharge with Municipal Effluent: Recharge Basins Soreq, Yavne 1, Yavne 2 & Yavne 3 : 2014*; Mekorot Water Company Limited, Central District, Dan Region Unit, 2014.

(6) Barkay Arbel Yoav; Kohen Efrat; Megidish Erez; Nadler Dorit; Amran Shlomi. *"Mey Ezor Dan" Agricultural Cooperative Water Society Ltd. Dan Region Wastewater Project Soreq Mechanical Biological Wastewater Treatment Plant Operation - 2021*; 2022.

(7) Elkayam, R.; Aharoni, A.; Vaizel-Ohayon, D.; Sued, O.; Katz, Y.; Negev, I.; Marano, R. B. M.; Cytryn, E.; Shtrasler, L.; Lev, O. Viral and Microbial Pathogens, Indicator Microorganisms, Microbial Source Tracking Indicators, and Antibiotic Resistance Genes in a Confined Managed Effluent Recharge System. *Journal of Environmental Engineering (United States)* **2018**, *144* (3). https://doi.org/10.1061/(ASCE)EE.1943-7870.0001334.

(8) TAHAL. *Geological Section Reports of the Israeli Costal Aquifer; Mekorot Water Company Report NO 732.*; 2001.

(9) Hunter, J. D. Matplotlib: A 2D Graphics Environment. *Computing in Science and Engineering*. 2007, pp 99–104. https://doi.org/10.1109/MCSE.2007.55.